%% file: Main-arXiv.tex
\documentclass[prx,twocolumn,amsmath,amssymb,floatfix,superscriptaddress]{revtex4-1}

\input{preamble}
\begin{document}

\title{Minimal Kitaev--transmon qubit based on double quantum dots}

\author{D. Michel Pino}
\affiliation{Instituto de Ciencia de Materiales de Madrid (ICMM), Consejo Superior de Investigaciones Científicas (CSIC), Sor Juana Inés de la Cruz 3, 28049 Madrid, Spain}
\author{Rubén Seoane Souto}
\affiliation{Instituto de Ciencia de Materiales de Madrid (ICMM), Consejo Superior de Investigaciones Científicas (CSIC), Sor Juana Inés de la Cruz 3, 28049 Madrid, Spain}
\author{Ramón Aguado}\email{ramon.aguado@csic.es}
\affiliation{Instituto de Ciencia de Materiales de Madrid (ICMM), Consejo Superior de Investigaciones Científicas (CSIC), Sor Juana Inés de la Cruz 3, 28049 Madrid, Spain}
\begin{abstract}
Minimal Kitaev chains composed of two semiconducting quantum dots coupled via a grounded superconductor have emerged as a promising platform to realize and study Majorana bound states (MBSs). We propose a hybrid qubit based on a Josephson junction between two such double quantum dots (DQDs) embedded in a superconducting qubit geometry. The qubit makes use of the $4\pi$-Josephson effect in the Kitaev junction to create a subspace based on the even/odd fermionic parities of the two DQD arrays hosting MBSs. Deep in the transmon regime, we demonstrate that by performing circuit QED spectroscopy on such hybrid Kitaev-Transmon "Kitmon" qubit one could observe distinct MBS features in perfect agreement with precise analytical predictions in terms of DQD parameters only. This agreement allows to extract the Majorana polarization in the junction from the microwave response.
\end{abstract}

\maketitle

\emph{Introduction}-- Majorana bound states (MBSs) appearing at the ends of one-dimensional topological superconductors~\cite{Leijnse_Review2012,Alicea_RPP2012,beenakker2013search,Aguado_Nuovo2017,BeenakkerReview_20,flensberg2021engineered,Marra_Review2022} feature non-abelian statistics that can be exploited for robust quantum information processing~\cite{Nayak_review}. Although early experiments showed signatures consistent with their presence, other states mainly originated from disorder can mimic their behavior, making it hard to distinguish between trivial and topological states~\cite{Prada-Review}.

Artificial Kitaev chains circumvent the inherent disorder issues that commonly appear in other platforms. In their minimal version, two quantum dots (QDs) couple via a narrow superconductor that allows for crossed Andreev reflection (CAR) and single-electron elastic co--tunneling (ECT) \cite{Leijnse,Sau_NatComm2012,PhysRevLett.129.267701,PhysRevB.106.L201404,Souto_arXiv2023,Bordin_PRX2023}. Minimal Kitaev chains can host localized MBSs when a so-called sweet spot is reached with equal CAR and ECT amplitudes. Although the states are not topologically protected, they share properties with their topological counterparts, including non-abelian statistics~\cite{tsintzis2023roadmap,Boross_2023}. Recent experiments have shown measurements consistent with predictions at the sweet spot regime~\cite{Dvir-Nature2023}, breaking a new ground for the investigation of MBSs and paving the way towards scaling a topologically-protected long chain and Majorana qubits \cite{10.1063/PT.3.4499} with QDs.
\begin{figure}[ht]
\centering
\includegraphics[width=\linewidth]{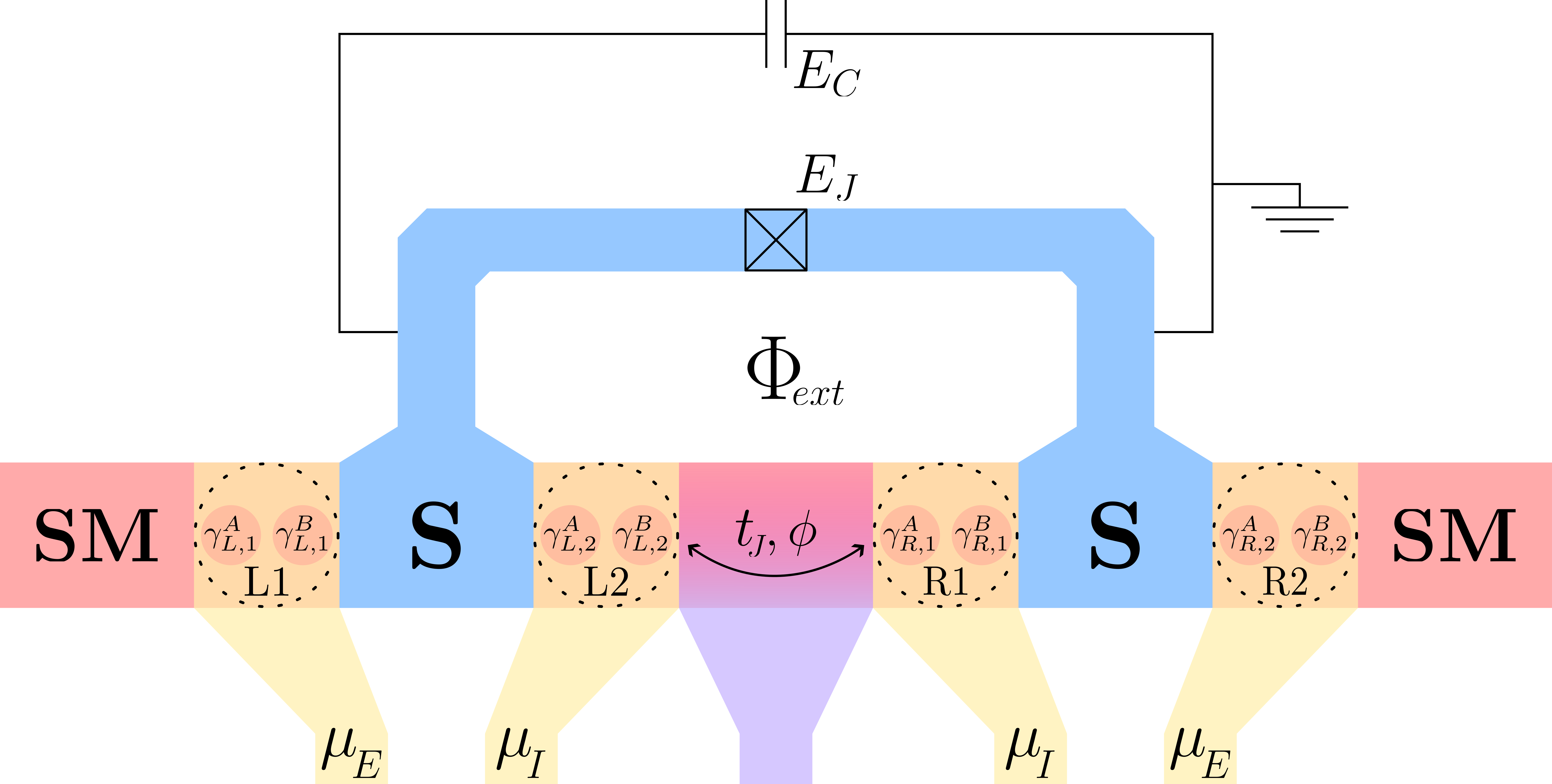}
\caption{{\bf Schematic illustration of the Kitaev-Transmon device}. A semiconductor (pink) can be gated (yellow) to create two minimal Kitaev chains (labeled as $\alpha=L,R$) comprising two quantum dots (labeled as $\beta=1,2$), connected via a middle superconductor (blue) and with chemical potentials $\mu_E$ and $\mu_I$, external and internal, respectively. Each quantum dot contains two Majorana states $\gamma_{\alpha,\beta}^A$ and $\gamma_{\alpha,\beta}^B$. The two Kitaev chains are connected through a weak link (hopping $t_J$, purple region) forming a minimal Majorana Josephson junction. This minimal Kitaev junction is connected to a transmon circuit, where the island, with charging energy $E_C$, is connected to ground by a SQUID formed by the parallel combination of the Kitaev junction and a reference Josephson junction $E_J$. The superconducting phase difference $\phi$ across the Kitaev junction is fixed by an externally applied magnetic flux $\Phi_{ext}$ applied through the SQUID loop.}
\label{fig:sketch}
\end{figure}

Expanding on this idea, we here propose a qubit based on a minimal Kitaev Josephson junction with four QDs and embedded in a superconducting qubit geometry, Fig. \ref{fig:sketch}. The Josephson potential of the QD array modifies the superconducting qubit Hamiltonian and splits the microwave (MW) transitions owing to the (nearly) degenerate fermionic parities of the Kitaev chains. Deep in the transmon limit, the qubit frequency can be analytically written in terms of QD parameters, Eq. (\ref{transition-eigenvalues}), in perfect agreement with full numerics (Fig. \ref{fig:panel_transmon}). This agreement allows to extract the Majorana polarization (MP) of the QD chain, Eq. (\ref{EM-MP}), a measure of the Majorana character of the ground states wavefunction~\cite{PhysRevB.106.L201404,Sedlmayr2015, Sedlmayr2016,Aksenov2020}, from the microwave response.

\emph{Model}--The minimal realization of a DQD-based Kitaev chain can be written as 
\begin{equation} \label{eq:Josephson_QD/chain_Hamiltonian}
H_{\mathrm{DQD}}  = -\sum_i\mu_{i} c_{i}^\dagger c_{i} 
 - tc_{1}^\dagger c_{2} + \Delta c_{1} c_{2}+\mbox{H.c.}\,,
\end{equation}
where $c_i^\dagger$ ($c_i$) denote creation (annihilation) operators on the $i\in 1,2$ quantum dot with a chemical potential $\mu_{i}$, while $t$ and $\Delta$ are the coupling strengths mediated by CAR and ECT processes across a middle superconducting segment, respectively \footnote{For the sake of simplicity, we assume in what follows that $t_\alpha$ and $\Delta_\alpha$ are parameters of the model, but we note in passing that both can be obtained from a microscopic description of the middle segments mediating the interdot couplings \cite{PhysRevLett.129.267701,PhysRevB.106.L201404}}.
Using this idea, a minimal Kitaev Josephson junction can be written as $H^{JJ}_{\mathrm{DQD}}=H_{\mathrm{DQD}}^L+H_{\mathrm{DQD}}^R+H_J$, where $H_{\mathrm{DQD}}^L$ and $H_{\mathrm{DQD}}^R$ are two left/right Kitaev chains based on Eq. (\ref{eq:Josephson_QD/chain_Hamiltonian}) and the Josephson coupling reads:
\begin{equation} \label{eq:Josephson_QD/chain_Hamiltonian2}
H_J=-t_J e^{i\phi/2}c_{L,2}^\dagger c_{R,1} + \mbox{H.c.} \;,
\end{equation}
with $\phi=\phi_R-\phi_L$ being the superconducting phase difference and $t_J$ the tunneling coupling between chains (see Fig. \ref{fig:sketch}).
The above model can be written in Bogoliubov--de Gennes (BdG) form as $H^{JJ}_{\mathrm{BdG}}= \frac{1}{2}\Psi^\dagger H^{JJ}_{\mathrm{DQD}}\Psi$, in terms of an eight-Majorana Nambu spinor
\begin{equation}\label{eq:Psi_state_basis}
\Psi = \left( \begin{matrix}
\gamma_{L,1}^A &
\gamma_{L,1}^B &
\gamma_{L,2}^A &
\gamma_{L,2}^B &
\gamma_{R,1}^A &
\gamma_{R,1}^B &
\gamma_{R,2}^A &
\gamma_{R,2}^B
\end{matrix} \right)^T\,.
\end{equation}
As we discuss below, the BdG model contains a standard Josephson coupling $\sim \cos\phi$ 
involving the "bulk" fermions together with a Majorana-mediated $4\pi$ Josephson effect
of order $\sim \cos\frac{\phi}{2}$. The latter involves coherent single-electron tunneling with a
characteristic energy scale $E_M$. From the perspective of circuit QED, previous papers have discussed how a Majorana junction in a transmon circuit
splits spectral lines corresponding to different fermionic parities owing to $E_M\neq 0$ \cite{Ginossar,Keselman,Yavilberg,Li2018,Avila,Avila2,Smith2020,Lupo2022}. In what follows, we discuss this physics in the context of the DQD minimal Kitaev Josephson junction and to analyse the novel aspects that arise when this promising new platform is integrated into a superconducting circuit.

\emph{Four Majoranas subspace}--A convenient way of gaining physical intuition is by projecting the above full model onto a low-energy subspace. The simplest approach,  widely used in previous literature \cite{PhysRevLett.108.257001,PhysRevB.86.140504,PhysRevB.97.041415,Cayao2018}, is to use a subspace spanned by just four MBSs: the two inner $\gamma_{L,2}^B$ and $\gamma_{R,1}^A$, and the two external $\gamma_{L,1}^A$ and $\gamma_{R,2}^B$. This results in an effective Josephson potential
\begin{equation} \label{eq:Majorana-transmon/Josephson_potential}
    V^{JJ}_{\mathrm{DQD}}(\phi) = E_M\cos\frac{\phi}{2}\sigma_x + E_M^S\sin\frac{\phi}{2}\sigma_y + \lambda\sigma_z,
\end{equation}
where $\sigma_i$ are Pauli matrices defined onto the pseudospin parity space spanned by $|0\rangle\equiv |n_L=0,\,n_R=0\rangle$ and $|1\rangle\equiv|n_L=1,\,n_R=1\rangle$, where $n_{L}=n_{L,1}+n_{L,2}$ and $n_{R}=n_{R,1}+n_{R,2}$ are the fermion occupations in the left/right segments of the junction. $E_M^S$ and $\lambda$ are due to additional inter and intra Majorana couplings \{$\gamma_{L,1}^A\leftrightarrow\gamma_{R,1}^A$, $\gamma_{L,2}^B\leftrightarrow\gamma_{R,2}^B$\} and \{$\gamma_{L,1}^A\leftrightarrow\gamma_{L,2}^B$, $\gamma_{R,1}^A\leftrightarrow\gamma_{R,2}^B$\}, respectively. In the symmetric case
$\mu_{L,1}=\mu_{R,2}=\mu_E$ and $\mu_{L,2}=\mu_{R,1}=\mu_I$, $E_{M}^S=\lambda=0$, which gives
\begin{equation}
 V^{JJ}_{\mathrm{DQD}}(\phi) =\frac{t_J}{2}\left[1-\frac{\mu_E^2}{(t+\Delta)^2}\right] \cos\frac{\phi}{2}\sigma_x\,.
\label{eq:Majorana-cupling-four}
\end{equation}

While being able to capture some of the phenomenology, including the $E_M$ renormalization with the external gates, this four--Majorana projection has important limitations.
Most importantly, detuning the chemical potentials $\mu_E$ and $\mu_I$ away from zero affects the localization of the MBSs which acquire some weight in "bulk" sites removed from the projection (for instance, a $\mu_E\neq 0$ induces weight of the order of $\sim {\mu_E\over t}$ in the inner dots~\cite{Leijnse}). This makes the four--Majorana projection insufficient to describe the physics of the DQD junction (for a full derivation of Eq.~\eqref{eq:Majorana-cupling-four} and a detailed discussion about the limitations of this projection, see Appendix I).

\begin{figure}[ht]
    \centering
    \includegraphics[width=\linewidth]{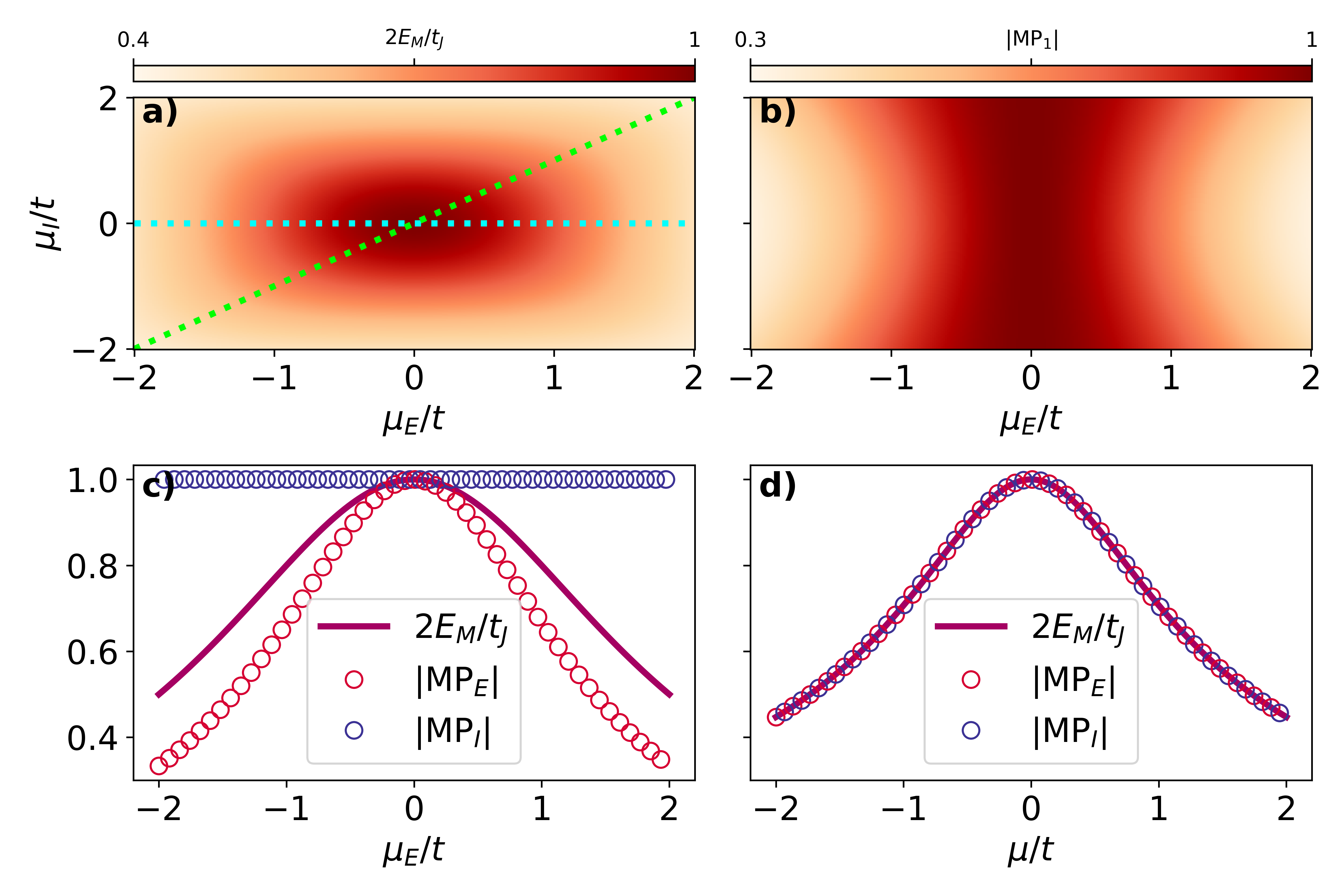}
    \caption{{\bf Majorana polarization and Majorana coupling}. (a) $2E_M/t_J$ and (b) $|\mathrm{MP}_{1}|$ as a function of $\mu_E$ and $\mu_I$. $2E_M/t_J$, $|\mathrm{MP}_{1}|$ and $|\mathrm{MP}_{2}|$ as a function of (c) $\mu_E$ with $\mu_I=0$ and (d) $\mu_E=\mu_I=\mu$ (blue and green dotted lines in panel a, respectively). $\Delta/t=1$ for all panels.}
\end{figure}
\begin{figure*}[ht]
    \centering
    \includegraphics[width=\linewidth]{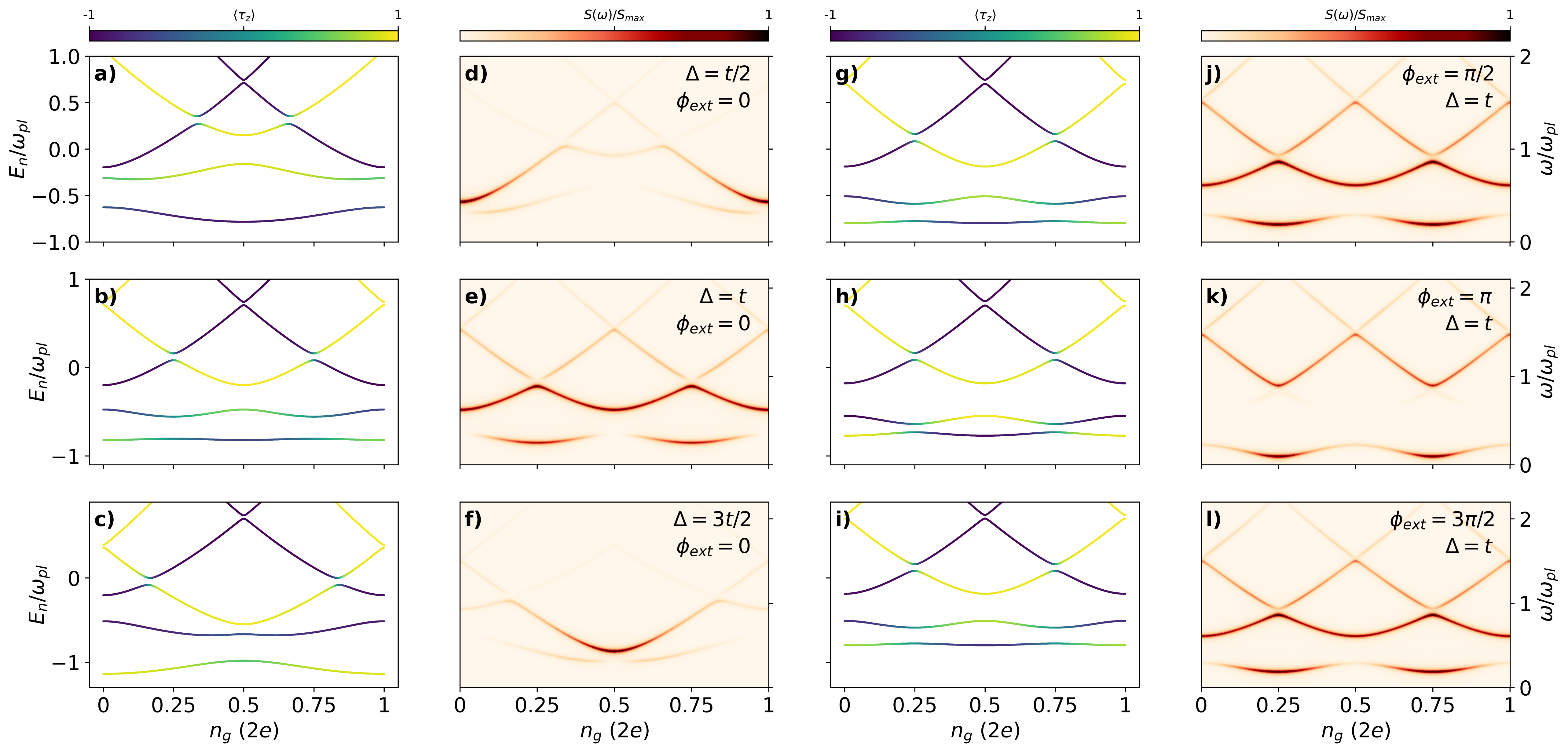}
    \caption{{\bf MW spectroscopy in charging regime}. Levels, parity texture $\langle \tau_z\rangle$ and $S(\omega)$ from the solutions of Eq. (\ref{eq:qubit_Hamiltonian}) against $n_g$ with $\mu_E=\mu_I=0$ for (a-f) $\Delta/t=0.5,\, 1,\, 1.5$ and $\phi_{ext}=0$; and (g-l) $\phi_{ext}=\pi/2, \pi, 3\pi/2$ and $\Delta=t$ (from top to bottom). $E_J/E_C=1$ and $t_J/t=1$ in all panels.}
    \label{fig:panel_CPB}
\end{figure*}
\emph{Beyond four Majoranas}--
To go beyond the previous projection and its limitations, we choose the subspace spanned by the two lowest--energy many--body eigenstates $\{|O_L^-,O_R^-\rangle,\,|E_L^-,E_R^-\rangle\}$ resulting from diagonalizing each isolated segment in the basis of occupation states $\{\ket{10},\,\ket{01},\,\ket{00},\,\ket{11}\}$. The diagonal Hamiltonian in the bipartite Hilbert space $\mathcal{H}_L\otimes\mathcal{H}_R$ can be represented on the basis of joint eigenstates $\{|i_L,j_R\rangle=|i_L\rangle\otimes|j_R\rangle\}$ with $i,j=O^\pm, E^\pm$ (see Appendix II):
\begin{equation}
\tilde{H}_L + \tilde{H}_R = (P_L^{-1} H_L P_L)\otimes\mathbb{I}_R + \mathbb{I}_L\otimes(P_R^{-1} H_R P_R), 
\end{equation}
where $P_\alpha$ is the change--of--basis matrix onto the eigenbasis of each chain with eigenenergies $\epsilon_{\alpha O}^\pm = -\mu_\alpha \pm \sqrt{t_\alpha^2+\delta_\alpha^2}$ and $\epsilon_{\alpha E}^\pm = -\mu_\alpha \pm \sqrt{\Delta_\alpha^2+\mu_\alpha^2}$, where we have defined $\mu_\alpha=(\mu_{\alpha,1}+\mu_{\alpha,2})/2$ and $\delta_\alpha=(\mu_{\alpha,1}-\mu_{\alpha,2})/2$. The off--diagonal Josephson term  $\tilde{H}_J$ can be easily represented on the joint--occupation basis $\{|n_{L,1},n_{L,2}\rangle\otimes|n_{R,1},n_{R,2}\rangle\}_{n_{\alpha,i}=0,1}$ and then projected onto the eigenbasis by the change--of--basis matrix $P_{LR}=P_L\otimes P_R$. Using this projection, the Josephson potential can be obtained analytically (see Appendix II). Specifically, for the mirror--symmetric case, $\mu_{L,1}=\mu_{R,2}=\mu_E$ and $\mu_{L,2}=\mu_{R,1}=\mu_I$ (external vs. internal), such that $\mu_L=\mu_R=(\mu_E+\mu_I)/2=\mu$ and $\delta_L=-\delta_R=(\mu_E-\mu_I)/2=\delta$, and considering $\Delta_L=\Delta_R$ and $t_L=t_R$, this Josephson potential reduces to a very compact form
\begin{equation} \label{eq:VJ_EO_IE}
V^{JJ}_{\mathrm{DQD}}(\phi)=\left(
\begin{matrix}
-2\mu -2\sqrt{t^2+\delta^2} & E_M\cos\frac{\phi}{2}
\\
E_M\cos\frac{\phi}{2} & -2\mu -2\sqrt{\Delta^2+\mu^2}
\end{matrix} \right)
\end{equation}
with
\begin{equation}
\label{EM-ManyBody}
E_M=\frac{t_J \Delta t}{2\sqrt{(t^2+\delta^2)(\Delta^2+\mu^2)}} \;.
\end{equation}

The diagonal terms in Eq.~\eqref{eq:VJ_EO_IE} originate from the MBSs overlapping within the same chain. Taking a series expansion up to leading order of $\mu$ and $\delta$, Eq. (\ref{EM-ManyBody}) reduces to $E_M$ in Eq. (\ref{eq:Majorana-cupling-four}) for $t=\Delta$ and $\mu=\delta$ ($\mu_I=0$).

\emph{Majorana polarization}--For $t_J=0$, the many body problem described above can be separated into two independent blocks of even ($\{|O_L^\pm,O_R^\pm\rangle$, $|E_L^\pm,E_R^\pm\rangle\}$) and odd ($\{|E_L^\pm,O_R^\pm\rangle,\,|E_L^\pm,O_R^\pm\rangle\}$) total parity, which leads to a two--fold degenerate spectrum. To determine whether these degeneracies are associated with MBSs, we use the Majorana polarization (MP) defined on the left Kitaev chain as
$\mathrm{MP}_{i}(O,E) =\frac{w_{i}^2 - z_{i}^2}{w_{i}^2 + z_{i}^2}$, 
with $ w_{i} = \brakettt{O}{c_{i}+c_{i}^\dagger}{E}$,
$z_{i} = \brakettt{O}{c_{i}-c_{i}^\dagger}{E}$ and $i\in 1,2$. For the left DQD, we take $\ket{E}=|O_L^-,O_R^-\rangle$, and $\ket{O}=|E_L^-,O_R^-\rangle$, this gives
\begin{equation}
\mathrm{MP}_{1/2}= \frac{t\Delta}{\pm\delta\mu - \sqrt{(t^2+\delta^2)(\Delta^2+\mu^2)}} \;,
\end{equation}
where we have omitted the left subscript for simplicity. A similar treatment can be performed for the right chain.

For $t=\Delta$, $|\mathrm{MP}_{1}|$ ($|\mathrm{MP}_{2}|$) is maximum when $\mu=\delta$ ($\mu=-\delta$), that is, when $\mu_{L,2}=0$ ($\mu_{L,1}=0$). Interestingly, by comparison with Eq. (\ref{EM-ManyBody}), when $\mu_{L,1}=\mu_{R,2}=\mu_E$ and $\mu_{L,2}=\mu_{R,1}=\mu_I$ ($\mu_L=\mu_R=\mu$ and $\delta_L=-\delta_R=\delta$), one can write:
\begin{equation}
\label{EM-MP}
\mathrm{MP}_{I/E} = \frac{-E_M}{\frac{t_J}{2} \pm \frac{\delta\mu}{t\Delta} E_M} \;.
\end{equation}
%with  $\beta=\{0\equiv I,\, 1\equiv E\}$. 
Note that for $\delta=0$ or $\mu=0$ ($\mu_E=\mu_I$ or $\mu_E=-\mu_I$, respectively), $\mathrm{MP}$ is equal on every QD and it is directly proportional to $E_M$. Therefore, Eq. (\ref{EM-MP}) directly relates the MP with $E_M$, which allows its direct measurement via MW spectroscopy as we discuss now.

\emph{Hybrid superconducting qubit model}--We now study a DQD-based
Majorana Josephson junction in a superconducting qubit geometry (namely a split junction shunted by a capacitor, with charging energy $E_C$, see Fig. \ref{fig:sketch}) described by the Hamiltonian:
\begin{equation} \label{eq:qubit_Hamiltonian}
H = 4E_C(\hat{n}-n_g)^2 -E_Jcos(\hat\phi)+V^{JJ}_{\mathrm{DQD}}(\hat\phi-\phi_{ext})\;.
\end{equation}
%with $E_C$ is the charging energy, $E_J$ the bulk Josephson coupling, 
Here, $\hat{n}=-i\frac{\partial}{\partial\hat\phi}$ is the Cooper-pair number operator, conjugate to the junction superconducting phase difference $\hat\phi$, and $n_g=Q_g/(2e)=V_g/(2eC_g)$ the gate--induced offset charge in the island (in units of Cooper pairs). The phase difference across the DQD Josephson junction can be controlled by the magnetic flux through the SQUID loop $\Phi_{ext}=\phi_{ext}\Phi_0/(2\pi)$, where
$\Phi_0 = h/2e$ is the superconducting flux quantum.
Using the solutions of (\ref{eq:qubit_Hamiltonian}) \footnote{
In practice, we solve the model as a tight--binding chain in charge space. Specifically, we divide the phase interval $\phi\in[0,2\pi)$ in $N$ steps, constructing a $2N\times 2N$ Hamiltonian matrix in tight--binding form. Then, we can move to its dual space of charge states $\{\ket{n},\ket{n+1/2}\}_{n=-N}^N$ and rewrite the tight--binding Hamiltonian in this basis by applying a Fourier transformation of the quantum phase operators (see Appendix V).}, the microwave (MW) absorption spectrum\footnote{For graphical purposes, we have convolved this quantity with a Cauchy--Lorentz distribution ($\gamma = 0.008$), which yields a finite--line broadening of the spectra.} of the superconducting island can be written in linear response as
$S(\omega) = \sum_k \left|\brakettt{k}{\hat{n}}{0}\right|^2 \delta(\omega-\omega_{0k}) \;$,
where the index $k$ orders the eigenstates of the system with increasing energies. This response measures the energy transitions $\omega_{0k} = \omega_k - \omega_0$ between the ground state $E_0 = \hbar \omega_0$ and the excited states $E_k = \hbar \omega_k$ and with a probability weighted by the matrix elements of ${\hat{n}}$.

Single--electron tunneling processes mediated by the off--diagonal terms of the DQD-based Josephson potential in Eq. (\ref{eq:VJ_EO_IE}) lead to very specific predictions in the spectrum that should be easily detected using standard circuit QED techniques. For example, crossing the sweet spot, while keeping $\mu_E=\mu_I=0$, from the ECT-dominated regime ($t>\Delta$, Fig. \ref{fig:panel_CPB}a) to the CAR-dominated regime ($t<\Delta$, Fig. \ref{fig:panel_CPB}c), changes the fermionic parity of the GS. This is reflected as an \emph{exact 1$e$ shift in $n_g$ in the MW spectra} (compare Figs. \ref{fig:panel_CPB}d and f). At the sweet spot for $t=\Delta$, the intraband coupling leads to maximally mixed parity states $\langle\tau_z\rangle=0$ with avoided crossings around $n_g=0.25$ and $n_g=0.75$, Fig. \ref{fig:panel_CPB}b. 
This results in an overall 1$e$-periodic MW spectrum with a strong low-frequency response near these gates, Fig. \ref{fig:panel_CPB}e.
Therefore, the intraband transition $\omega_{01}$ is a direct indication of a $E_M\neq 0$ in the spectrum. 

\emph{Kitaev-Transmon regime}--A way to check that the low-frequency MW transitions $\omega_{01}$ near $n_g=0.25$ and $n_g=0.75$ are indeed due to parity mixing mediated by MBSs in the DQD junction, instead of quasiparticle poisoning \cite{SchreierPRB08,KringhojPRL20,BargerbosPRL20}, is to prove that they can be tuned by $\phi_{ext}$, and reach a minimum value at $\phi_{ext}=\pi$, Figs. \ref{fig:panel_CPB}j-l. Note however, that owing to quantum phase fluctuations, the Josephson potential $V^{JJ}_{\mathrm{DQD}}$ in Eq. (\ref{eq:qubit_Hamiltonian}) depends on a phase drop which
deviates from the external phase imposed by the circuit, hence resulting in a residual splitting at $n_g=0.25$ which does not close completely at $\phi_{ext}=\pi$. This effect is shown in Fig. \ref{fig:panel_transmon}a, where we plot the full $\phi_{ext}$ dependence corresponding to the MW spectra of Figs. \ref{fig:panel_CPB}j-l at fixed $n_g=0.25$. Interestingly, parity changes due to Majorana physics are already evident as a spectral hole near $\phi_{ext}=\pi$ in the transition $\omega_{02}$. By tracing such spectral hole in $\omega_{02}$ (or, equivalently the appearance of the transition $\omega_{03}$) we can identify when a true energy crossing occurs in the system as a function of increasing $E_J/E_C$ ratios, Figs. \ref{fig:panel_transmon}b,c.
\begin{figure}[ht]
    \centering
    \includegraphics[width=\linewidth]{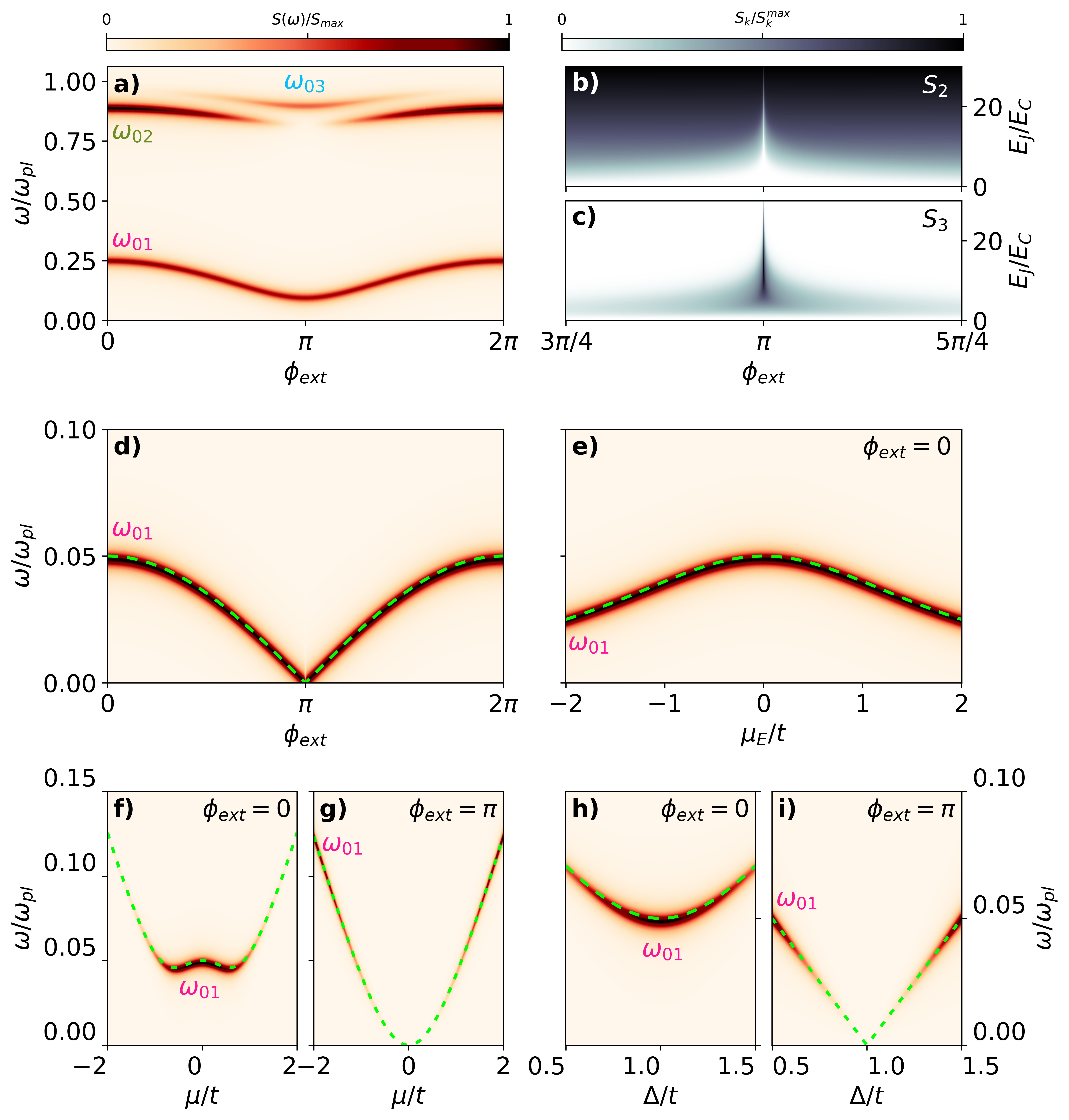}
    \caption{{\bf Kitaev-transmon qubit spectroscopy}. (a) Full phase dependence of the MW absorption spectrum of Fig. \ref{fig:panel_CPB}g-l at $n_g=0.25$. (b-c) Spectral weights for transitions $\omega_{02}$ ($S_2$) and $\omega_{03}$ ($S_3$) as a function of $\phi_{ext}$ and $E_J/E_C$ at the sweet spot ($\Delta=t$, $\mu_E=\mu_I=0$). (d-g) MW absorption spectra as a function of (d) $\phi_{ext}$ at the sweet spot; (e) $\mu_E$ with $\mu_I=0$ and $\Delta=t$ and $\phi_{ext}=0$; (f-g) $\mu_E=\mu_I=\mu$ with $\Delta=t$ and $\phi_{ext}=0,\pi$; and (h-i) $\Delta/t$ with $\mu_E=\mu_I=0$ and $\phi_{ext}=0,\pi$. Green dashed lines correspond to the analytical qubit frequency $\omega_{KiT}$ in Eq. (\ref{transition-eigenvalues}). For panels (d-g) we have fixed a ratio $E_J/E_C=50$. $t_J/t=1$ for all panels.}
    \label{fig:panel_transmon}
\end{figure}
While, generally, an analytical expression of the energy splitting at $n_g=0.25$ would require knowing the explicit form of the qubit wave functions, the deep transmon regime with $E_J/E_C\gg 1$ allows us to approximate these eigenfunctions to two coupled (parity--defined) harmonic--oscillator states sharpened around $\phi_{ext}$. In this regime, the Kitmon qubit frequency $\omega_{KiT}\equiv\omega_{01}$ can be written as
\begin{equation}
\label{transition-eigenvalues}
\omega_{KiT} \approx 2\sqrt{(\sqrt{t^2+\delta^2} - \sqrt{\Delta^2+\mu^2})^2 + E_M^2\cos^2\frac{\phi_{ext}}{2}}
\end{equation}
(A detailed check of the validity of Eq. (\ref{transition-eigenvalues}) for increasing values of $E_J/E_C$ ratios can be found in Appendix IV). When $t=\Delta$ and $\delta=\pm\mu$ ($\mu_I=0$ or $\mu_E=0$), the qubit frequency is directly proportional to $E_M$,
\begin{equation}
\label{Pino resonance}
   \omega_{KiT} \approx 2 E_M \cos\frac{\phi_{ext}}{2}=\frac{t_J}{1+(\mu_E/\Delta)^2}\cos\frac{\phi_{ext}}{2}.
\end{equation}

A direct comparison between the full numerics and Eq. (\ref{transition-eigenvalues}) against different parameters of the junction, Figs. \ref{fig:panel_transmon}d-i, demonstrates an almost perfect agreement. Therefore, MW measurements like the ones discussed here should allow to check our predictions, e.g., the resonant behavior against $\mu_E$ in Eq. (\ref{Pino resonance}), see Fig. \ref{fig:panel_transmon}e. More importantly, a measurement like the one shown in Figs. \ref{fig:panel_transmon}f and g (namely, $\omega_{KiT}$ versus $\mu=\mu_E=\mu_I$, hence $\delta=0$) would allow to directly extract $E_M$ and hence determine the MP polarization of the junction via Eq. (\ref{EM-MP}).

In conclusion, we have proposed a minimal Kitaev-Transmon qubit based on a QD Josephson junction array embedded in a superconducting circuit. Deep in the transmon regime with $E_J/E_C\gg 1$ we have found an analytical expression for the qubit frequency, Eq. (\ref{transition-eigenvalues}), that allows to obtain very precise predictions of its evolution against QD parameters, Fig. \ref{fig:panel_transmon}, and to extract the Majorana polarization. The precise predictions in terms of analytics would allow to experimentally distinguish the physics discussed here from either quasiparticle poisoning or 4$\pi$ phase slips due to QD resonances \cite{Vakhtel23}. This novel qubit architecture is a natural extension of the recent experimental implementations of nanowire-based double island devices~\cite{Zanten_NatPhys2020}, gatemons~\cite{gatemon1,gatemon2,Sabonis_PRL2020,Huo_2023} and Andreev spin qubits \cite{Hays2021}, although free from the uncertainties originated from disorder.
Most importantly, QD-based Josephson junctions embedded in a transmon circuit have recently been implemented experimentally  \cite{KringhojPRL20,BargerbosPRL20,PRXQuantum.3.030311}. In the strong Coulomb Blockade regime, they have been used to show spin-split MW transition lines \cite{Bargerbos2022spectroscopy} forming a QD-based superconducting spin qubit coherently coupled to a transmon \cite{Pita-Vidal-NaturePhys23}. In this context, our DQD proposal could be seen as a minimal Majorana-based non-local parity pseudospin, Eq.~\eqref{eq:VJ_EO_IE}, coupled to a transmon. All this experimental progress, together with the recent demonstration of poisoning times of the order of milliseconds~\cite{Hinderling_arXiv2023} and quasiparticle trapping engineering \cite{Gerbold19,NguyenPRB23,uilhoorn2021quasiparticle}, make the physics discussed here within reach with superconductor-semiconductor hybrid devices \footnote{Two-tone spectroscopy measurements used to detect the MW transitions described here are typically integrated in time scales of the order of tens of milliseconds, see e.g. \cite{BargerbosPRL20}}. 
\begin{acknowledgments}
We acknowledge the support of the Spanish Ministry of Science through Grants PID2021- 125343NB-I00 and TED2021-130292B-C43 funded by MCIN/AEI/10.13039/501100011033, "ERDF A way of making Europe", the Spanish CM “Talento Program” (project No. 2022-T1/IND-24070), and European Union NextGenerationEU/PRTR. Support by the CSIC Interdisciplinary Thematic Platform (PTI+) on Quantum Technologies (PTI-QTEP+) is also acknowledged.
\end{acknowledgments}

\bibliographystyle{ieeetr}

\bibliography{bibliography}

\clearpage
\onecolumngrid
\setcounter{figure}{0}
\setcounter{equation}{0}

\begin{center}
\large{\bf Supplemental Material\\}
\end{center}

\renewcommand\thefigure{S\arabic{figure}}
\renewcommand{\tablename}{Table.~S}
\renewcommand{\thetable}{\arabic{table}}

\setcounter{equation}{0}
\setcounter{figure}{0}

\renewcommand{\theequation}{S.\arabic{equation}}
\renewcommand{\thefigure}{S.\arabic{figure}}

\section{Four Majoranas subspace}
\subsection{Effective low--energy projection} \label{sec:supplemental/effective-four-Majoranas}

In order to derive a quantitative low--energy description of our system, we project the full Hamiltonian $H^{JJ}_{\mathrm{DQD}}$ --Eqs. (1) and (2) of the main text-- onto the fermionic parity subspace that forms the superconducting qubit. This procedure considers both standard Josephson events due to Cooper pair tunneling, as well as anomalous Majorana--mediated events, where a single electron is transferred across the junction. Hence, we can distinguish two contributions of the Josephson potential, $V_J = V_J^{\mathrm{bulk}} + V_{\mathrm{DQD}}^{JJ}$. The first one takes into account the bulk contribution of the Bogoliubov--de Gennes (BdG) levels above the gap to the ground state energy, which we just assume to be of the standard form $V_J^{\mathrm{bulk}}(\phi)=-E_J\cos\phi$. The second contribution corresponds to the subgap sector, and it can be expressed as the projection onto a fermionic parity basis of an effective model of four Majorana operators, $\gamma_{L,1}^A,\gamma_{L,2}^B\in L$ and $\gamma_{R,1}^A,\gamma_{R,2}^B\in R$, corresponding to the end modes of both chains. Its effective Hamiltonian takes the general BdG form
\begin{equation} \label{eq:supplemental/effective_Hamiltonian_Majorana}
H_\gamma = \frac{i}{2}\left(\begin{matrix}
\gamma_{L,1}^A & \gamma_{L,2}^B & \gamma_{R,1}^A & \gamma_{R,2}^B
\end{matrix}\right) \left(\begin{matrix}
0 & \lambda_{L1,L2} & \lambda_{L1,R1} & \lambda_{L1,R2}
\\
-\lambda_{L1,L2} & 0 & \lambda_{L2,R1} & \lambda_{L2,R2}
\\
-\lambda_{L1,R1} & -\lambda_{L2,R1} & 0 & \lambda_{R1,R2}
\\
-\lambda_{L1,R2} & -\lambda_{L2,R2} & -\lambda_{R1,R2} & 0
\end{matrix}\right) \left(\begin{matrix}
\gamma_{L,1}^A \\ \gamma_{L,2}^B \\ \gamma_{R,1}^A \\ \gamma_{R,2}^B
\end{matrix}\right) \;.
\end{equation}

Our objective is now to relate $H^{JJ}_{\mathrm{DQD}}$ to this general effective model of four Majoranas $H_\gamma$ to obtain an explicit expression of its coefficients. Thus, we project the BdG form of the former, $H^{JJ}_{\mathrm{BdG}}$ --Eqs. (1) and (2) of the main text using the Majorana spinor in Eq. (3) of the main text--, onto the low--energy subspace of Majorana operators. In order to do that, we define a basis of fermionic operators
\begin{equation} \label{eq:supplemental/Majorana_decomposition_QD}
c_\alpha = \frac{1}{\sqrt{2}}(\gamma_{\alpha,1}^A + i\gamma_{\alpha,2}^B) \quad,\quad c_\alpha^\dagger = \frac{1}{\sqrt{2}}(\gamma_{\alpha,1}^A -i\gamma_{\alpha,2}^B)\;,
\end{equation}
and we compute the matrix elements of the resolvent of $H^{JJ}_{\mathrm{BdG}}$,
\begin{equation}
G(\omega) = [(\omega + i\,\epsilon) \mathbb{I} - H^{JJ}_{\mathrm{BdG}}]^{-1} \;,\quad \epsilon\to 0^+\;,
\end{equation}
at $\omega=0$ on the $\psi^0 = (c_L,c_R,c_L^\dagger,c_R^\dagger)^T_0$ state basis. The procedure is as follows: first of all, we calculate $G(\omega)$ by inverting the matrix $(\omega+i\,\epsilon)\mathbb{I}-H^{JJ}_{\mathrm{BdG}}$ written on the state basis of the whole system,
\begin{equation} \label{eq:supplemental/Majorana_Nambu-basis}
    \Psi = \left( \begin{matrix}
        \gamma_{L,1}^A & \gamma_{L,1}^B &
        \gamma_{L,2}^A &
        \gamma_{L,2}^B & \gamma_{R,1}^A & \gamma_{R,1}^B & \gamma_{R,2}^A & \gamma_{R,2}^B &
    \end{matrix} \right)^T \quad,\quad \Psi^\dagger = \Psi^T \;.
\end{equation}

Then, we evaluate this resolvent matrix at $\omega=0$ and we project it onto the $\psi^0$ basis, expressed in terms of $\Psi$ states as
\begin{equation}
\begin{aligned}
\left(\begin{matrix}
1 & 0 & 0 & 0
\end{matrix}\right)^T_0 & \equiv \frac{1}{\sqrt{2}}\left(\begin{matrix}
1 & 0 & 0 & i & 0 & 0 & 0 & 0
\end{matrix}\right)^T
\quad &,\quad
\left(\begin{matrix}
0 & 1 & 0 & 0
\end{matrix}\right)^T_0 & \equiv \frac{1}{\sqrt{2}}\left(\begin{matrix}
0 & 0 & 0 & 0 & 1 & 0 & 0 & i
\end{matrix}\right)^T
\\
\left(\begin{matrix}
0 & 0 & 1 & 0
\end{matrix}\right)^T_0 & \equiv \frac{1}{\sqrt{2}}\left(\begin{matrix}
1 & 0 & 0 & -i & 0 & 0 & 0 & 0
\end{matrix}\right)^T
\quad &,\quad
\left(\begin{matrix}
0 & 0 & 0 & 1
\end{matrix}\right)^T_0 & \equiv \frac{1}{\sqrt{2}}\left(\begin{matrix}
0 & 0 & 0 & 0 & 1 & 0 & 0 & -i
\end{matrix}\right)^T \;.
\end{aligned}
\end{equation}

This gives rise to a $4\times 4$ matrix
\begin{equation}
(\mathcal{H}_0^{-1})_{ij} = \brakettt{\psi^0_i}{G(\omega=0)}{\psi^0_j}\;,
\end{equation}
whose inverse
\begin{equation}
H_0 = \frac{1}{2}\sum_{i,j}\psi^{0\dagger}_i(\mathcal{H}_0)_{ij}\psi^0_j\;,
\end{equation}
is the projection of $H^{JJ}_{\mathrm{DQD}}$ onto the subspace of low--energy fermions. Finally, a simple change of basis $\psi^0\to\psi^\gamma=(\gamma_{L,1}^A,\gamma_{L,2}^B,\gamma_{R,1}^A,\gamma_{R,2}^B)_\gamma^T$ will allow us to indentify this matrix $\mathcal{H}_0$ with the effective sub--gap Hamiltonian (\ref{eq:supplemental/effective_Hamiltonian_Majorana}). Indeed, writing the $\psi^\gamma$ basis states in terms of $\psi^0$ components,
\begin{equation}
\begin{aligned}
\left(\begin{matrix}
1 & 0 & 0 & 0
\end{matrix}\right)^T_\gamma \equiv \frac{1}{\sqrt{2}}\left(\begin{matrix}
1 & 0 & 1 & 0
\end{matrix}\right)^T_0
\quad &,\quad
\left(\begin{matrix}
0 & 1 & 0 & 0
\end{matrix}\right)^T_\gamma \equiv \frac{1}{\sqrt{2}}\left(\begin{matrix}
-i & 0 & i & 0
\end{matrix}\right)^T_0\;,
\\
\left(\begin{matrix}
0 & 0 & 1 & 0
\end{matrix}\right)^T_\gamma \equiv \frac{1}{\sqrt{2}}\left(\begin{matrix}
0 & 1 & 0 & 1
\end{matrix}\right)^T_0
\quad &,\quad
\left(\begin{matrix}
0 & 0 & 0 & 1
\end{matrix}\right)^T_\gamma \equiv \frac{1}{\sqrt{2}}\left(\begin{matrix}
0 & -i & 0 & i
\end{matrix}\right)^T_0 \;,
\end{aligned}
\end{equation}
we can express the Hamiltonian $\mathcal{H}_0$ in this new basis as
\begin{equation}
(\mathcal{H}_\gamma)_{ij} = \langle\psi^\gamma_i|\mathcal{H}_0|\psi^\gamma_j\rangle \;,
\end{equation}
which yields
\begin{equation}
\label{Effective Josephson4Majo}
\begin{aligned}
& H_\gamma = \frac{1}{2}\sum_{ij}\psi^{\gamma\dagger}_i (\mathcal{H}_\gamma)_{ij}\psi^\gamma_j =
\\
& \frac{i}{2}\psi^{\gamma\dagger}\left(\begin{matrix}
0 & \frac{\mu_{L,1}\mu_{L,2} - (t_L+\Delta_L)(t_L-\Delta_L)}{t_L+\Delta_L} & -\frac{t_J \mu_{L,1} \sin\frac{\phi}{2}}{t_L+\Delta_L} & -\frac{t_J\mu_{L,1}\mu_{R,2}\cos\frac{\phi}{2}}{(t_L+\Delta_L)(t_R+\Delta_R)}
\\
\frac{(t_L+\Delta_L)(t_L-\Delta_L) - \mu_{L,1}\mu_{L,2}}{t_L+\Delta_L} & 0 & t_J\cos\frac{\phi}{2} & -\frac{t_J\mu_{R,2}\sin\frac{\phi}{2}}{t_R+\Delta_R}
\\
\frac{t_J \mu_{L,1} \sin\frac{\phi}{2}}{t_L+\Delta_L} & -t_J\cos\frac{\phi}{2} & 0 & \frac{\mu_{R,1}\mu_{R,2} - (t_R+\Delta_R)(t_R-\Delta_R)}{t_R+\Delta_R}
\\
\frac{t_J\mu_{L,1}\mu_{R,2}\cos\frac{\phi}{2}}{(t_L+\Delta_L)(t_R+\Delta_R)} & \frac{t_J\mu_{R,2}\sin\frac{\phi}{2}}{t_R+\Delta_R} & \frac{(t_R+\Delta_R)(t_R-\Delta_R) - \mu_{R,1}\mu_{R,2}}{t_R+\Delta_R} & 0
\end{matrix}\right)\psi^\gamma \;.
\end{aligned}
\end{equation}

Therefore, we can identify each element of this matrix with one coefficient $\lambda_{\alpha\beta}$ of Eq. (\ref{eq:supplemental/effective_Hamiltonian_Majorana}). It should be noted that this identification is an approximation; also, the separation between bulk and subgap contributions is only well-defined if the subgap modes are well-detached from the quasicontinuum.

\subsection{Comparison between eight and four Majoranas}

\begin{figure*}[ht]
\centering
    \includegraphics[width=0.8\linewidth]{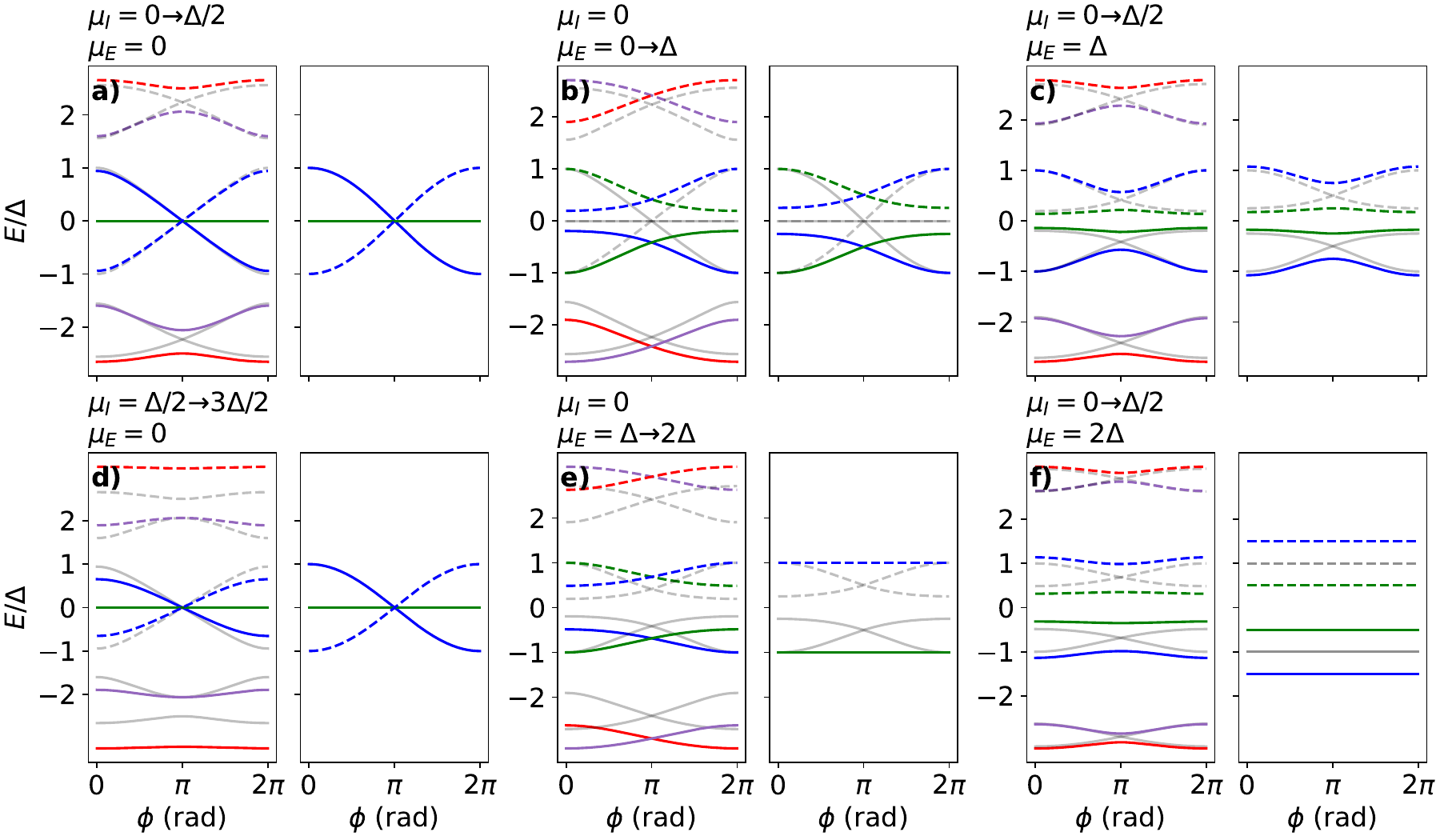}
\caption{Evolution of the energy spectrum as a function of $\phi$ for the parameter trajectory indicated in each panel. In each case, the leftmost panels correspond to the BdG form of the full Hamiltonian --Eqs. (1) and (2) using the Majorana spinor (3) of the main text-- and the rightmost panels to the four Majoranas projection --Eq. \ref{Effective Josephson4Majo}--. Gray/colored levels denote the beginning/end of each trajectory. We have fixed $t_J=t=\Delta$ for every panel.}
\label{fig:supplemental/spectrum-phi-traj}
\end{figure*}

Since our main objective is to study the physics of a superconducting qubit modified by the presence of the DQD Josephson junction, we first check the limitations of the effective Josephson potential obtained previously. At this level, it is enough to compare results from the projected potential in Eq. (\ref {Effective Josephson4Majo}) with the phase-dependent energy spectrum $E(\phi)$ of the BdG form of the full Hamiltonian $H^{JJ}_{\mathrm{BdG}}$ before any projection, Fig. \ref{fig:supplemental/spectrum-phi-traj}. At the sweet spot ($\Delta=t$, $\mu_E=\mu_I=0$, Fig. \ref{fig:supplemental/spectrum-phi-traj}a), the subgap spectrum shows a $4\pi$ Josephson effect indicating the presence of Majorana zero modes (thin grey lines). This spectrum originates from the fusion (energy splitting) of the inner MBSs living in the junction $\gamma_{L,2}^B$ and  $\gamma_{R,1}^A$ (which is maximum at $\phi=0,2\pi$), but without breaking the degeneracy point at $\phi=\pi$. Moreover, two states remain at zero energy for all phases, corresponding to the Majorana states $\gamma_{L,1}^A$ and  $\gamma_{R,2}^B$ living in the outermost quantum dots. In this regime, both the full solution (left panel) and the four MBSs projection (right panel) coincide. Of course, the latter does not capture the bulk solutions that disperse with phase near $2\Delta=2t$. Deviations from the sweet spot by changing the internal chemical potential $\mu_I\neq 0$ do not affect the low energy spectrum but open gaps in the bulk (colored lines). When moving away from the sweet spot by tuning the external chemical potentials $\mu_E\neq 0$, while keeping $\mu_I=0$, the spectrum remains $4\pi$--periodic. In this case, the low-energy states are lifted away from zero energy, Fig.~\ref{fig:supplemental/spectrum-phi-traj}b blue/green colored lines, resulting in a characteristic diamond-like shape. The crossings forming the diamonds become avoided crossings for $\mu_I\neq 0$ and $\mu_E\neq 0$, Fig.~\ref{fig:supplemental/spectrum-phi-traj}c, which also splits the crossings of the bulk bands near $\phi=\pi$, giving an overall $2\pi$-periodic spectrum. In contrast, a zero-energy state persists for $\mu_E=0$ and independently from $\mu_I$, even at large values, Fig.~\ref{fig:supplemental/spectrum-phi-traj}d, corresponding to the Majorana states of the outermost dots having zero weight in the inner ones. In this regime, the effect of detuning $\mu_I$ away from the sweet spot only affects the localization of the inner Majorana state, decreasing the splitting between the blue states, and resulting in a robust $4\pi$-periodic spectrum.

In all the cases described above, the approximation derived in Eq.~\eqref{Effective Josephson4Majo} using four Majorana states describes well the low-energy states of the system close to the sweet spot. In contrast, this approximation largely deviates from the results of the full Hamiltonian for sufficiently large $\mu_E\gtrsim\Delta$ and irrespective of $\mu_I$, Figs.~\ref{fig:supplemental/spectrum-phi-traj}e--f. In this regime, the bulk solutions that appear at $E\sim2\Delta$ at the sweet spot, hybridize with the low-energy states, renormalizing their energy and strongly affecting their dispersion against phase. Therefore, the low-energy states cannot be described by only four Majorana states (one per dot).

\begin{figure*}[ht]
\centering
    \subfloat{
        \includegraphics[width=0.5\linewidth]{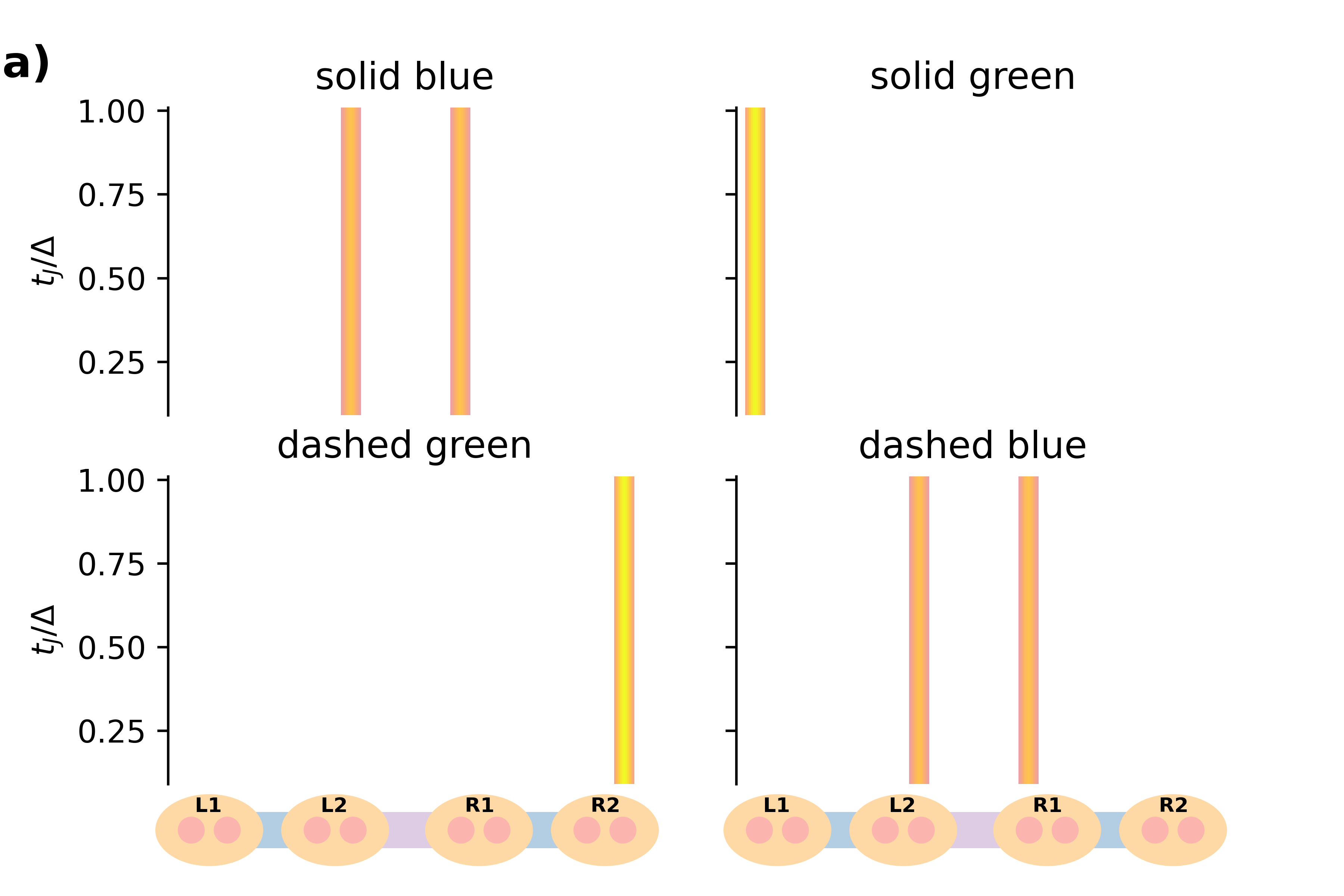}
    }
    \subfloat{
        \includegraphics[width=0.5\linewidth]{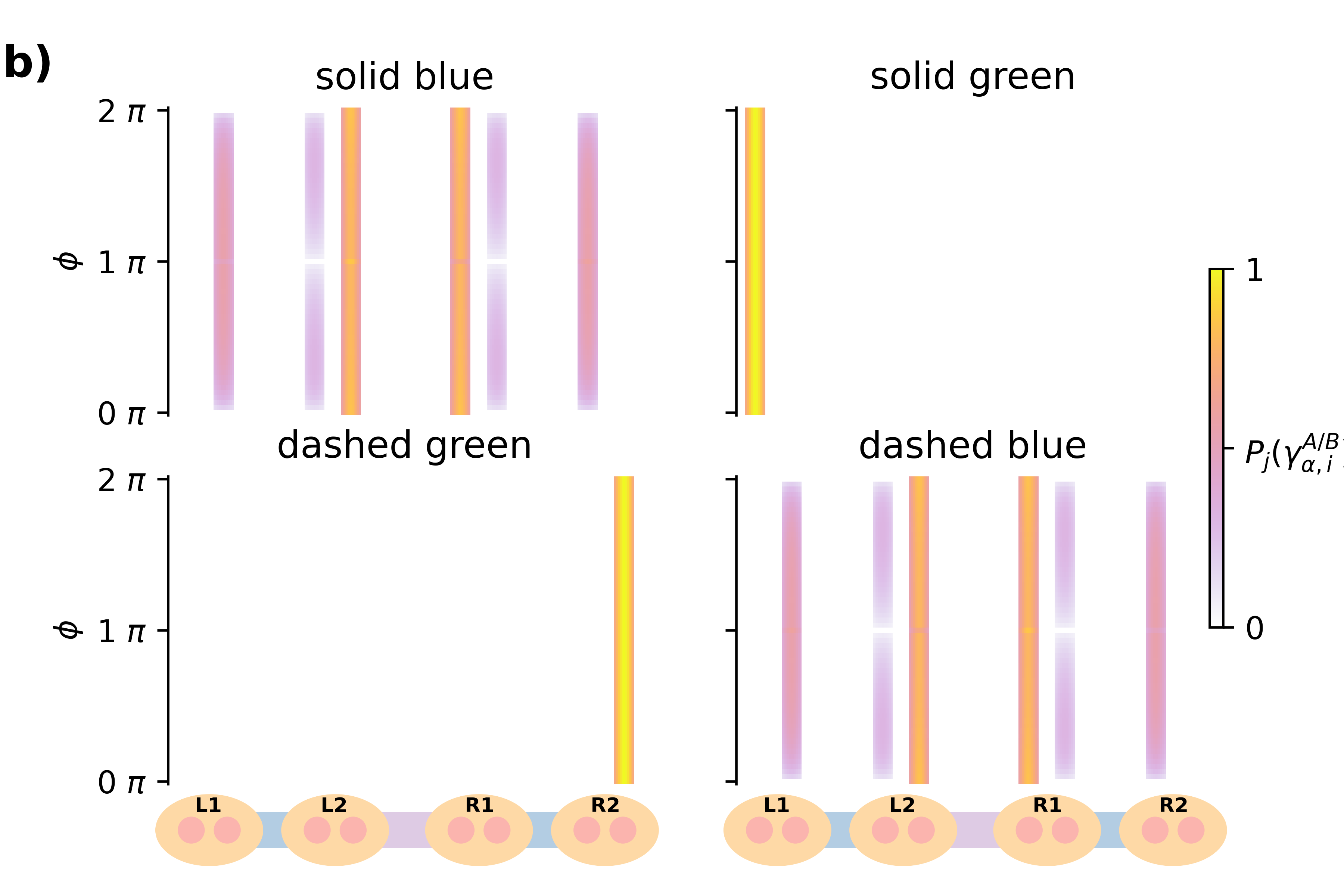}
    }\\
    \subfloat{
        \includegraphics[width=0.5\linewidth]{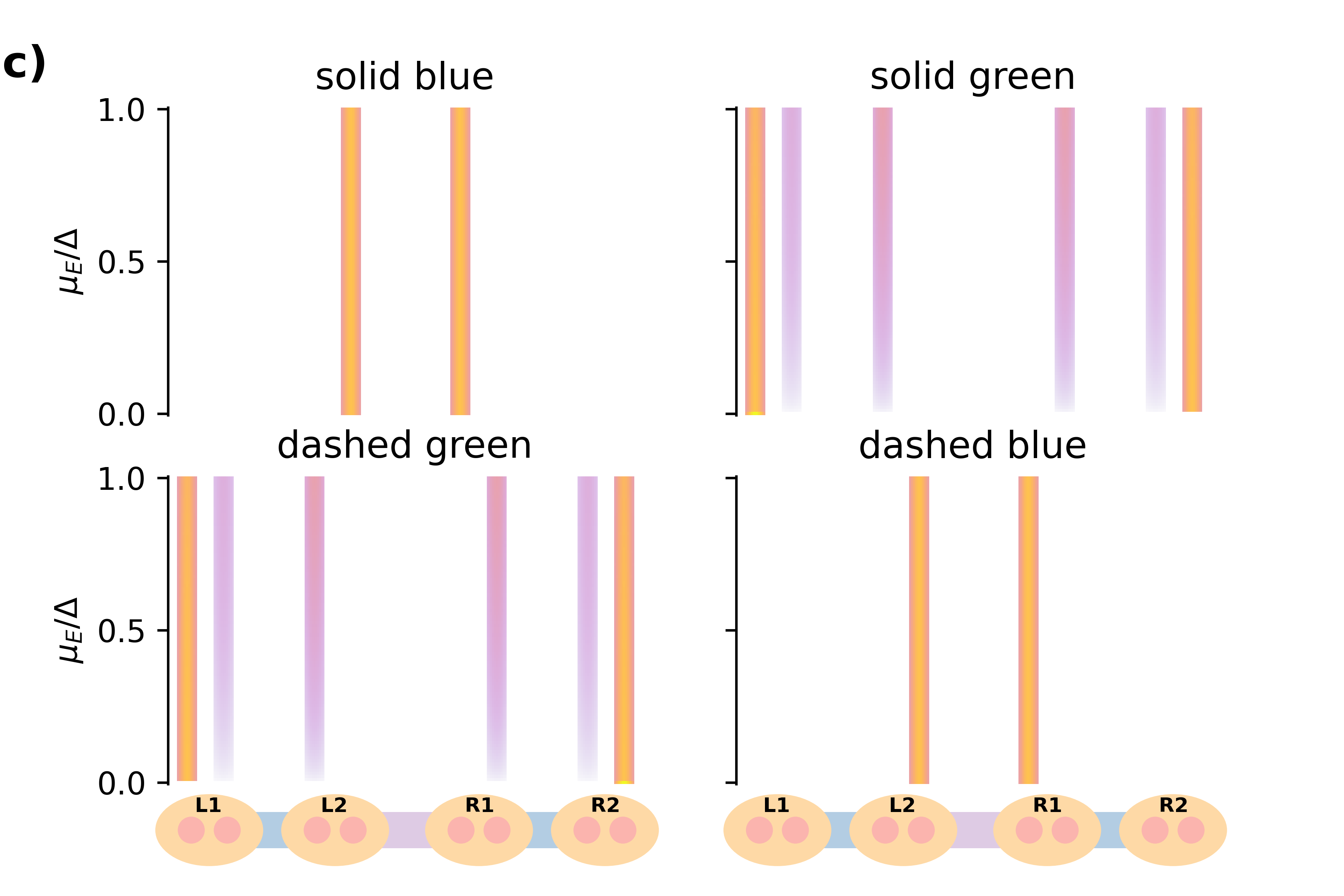}
    }
    \subfloat{
        \includegraphics[width=0.5\linewidth]{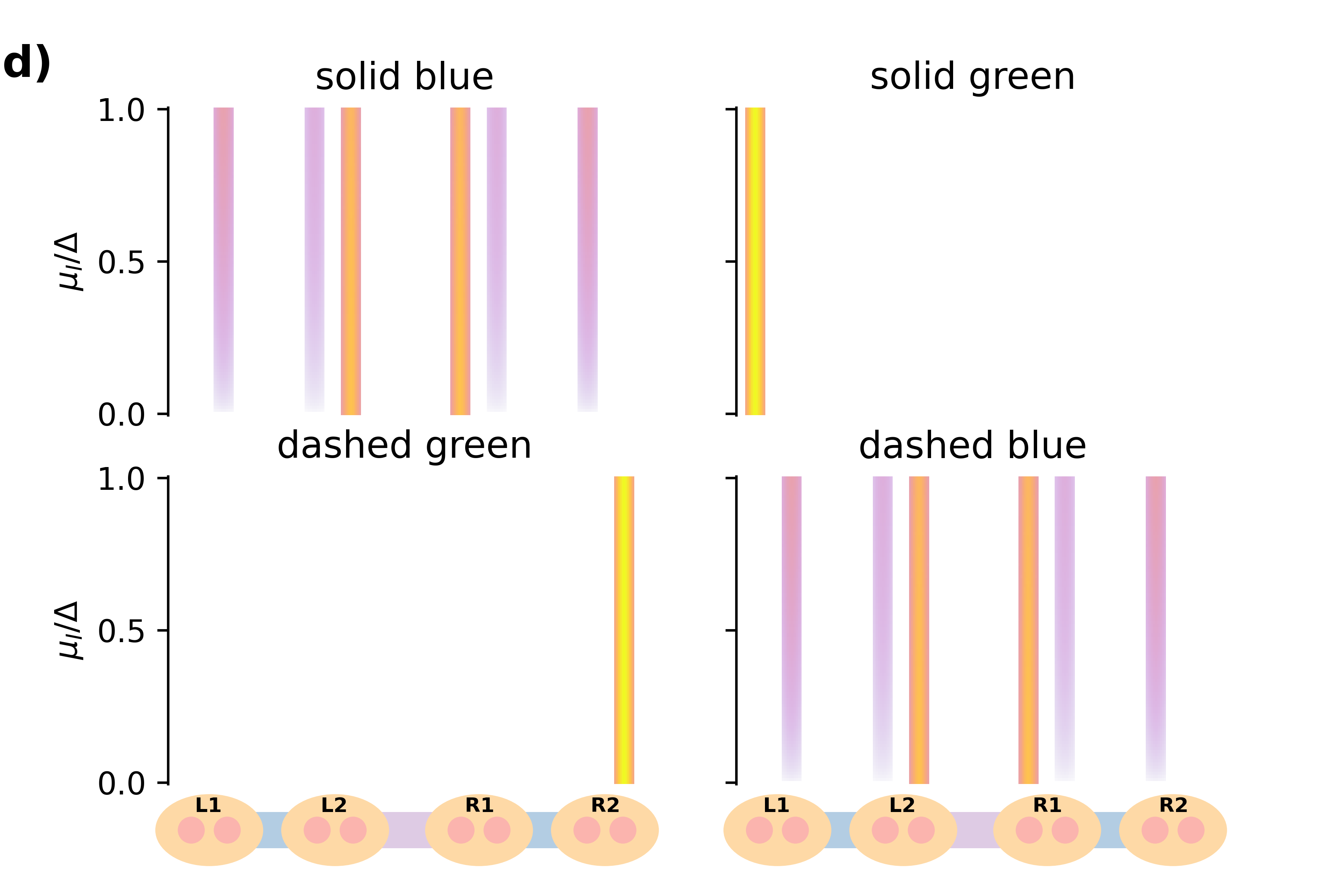}
    }
\caption{Evolution of the space distribution of sub--gap states as a function of (a) $t_J$ with $\mu_E=\mu_I=0$ and $\phi=0$; (b) $\phi$ with $\mu_E=\mu_I=0$; (c) $\mu_E$ with $\mu_I=0$ and $\phi=0$; and (d) $\mu_I$ with $\mu_E=0$ and $\phi=0$. We have fixed $\Delta=t=t_J$ for all panels, and subtitles refer to each eigenstate plotted in Fig. \ref{fig:supplemental/spectrum-phi-traj}.}
\label{fig:supplemental/localization_modes}
\end{figure*}

We demonstrate the importance of considering all the Majorana states in every dot by calculating the real space--resolved distribution of the wave functions, taken as the probability  $P_j(\gamma_{\alpha,i}^{A/B}) = \langle \psi_j | \Psi_{\alpha,i}^{A/B}\rangle\langle \Psi_{\alpha,i}^{A/B}| \psi_j\rangle$ of the eigenstate $\ket{\psi_j}$ of $H^{JJ}_{\mathrm{BdG}}$ on each mode $\gamma_{\alpha,i}^{A/B}$, represented in the Majorana basis (\ref{eq:supplemental/Majorana_Nambu-basis}). Here indices $i=1,2$ and $\alpha=L,R$ denote the sites of each chain, whereas $j=\text{green}, \text{blue}$ labels the different levels that appear in Fig. \ref{fig:supplemental/spectrum-phi-traj}. As we can see in Fig. \ref{fig:supplemental/localization_modes}, at the sweet spot the outermost Majoranas are pinned to zero energy (green states in Fig.~\ref{fig:supplemental/spectrum-phi-traj}), whereas (oscillating) blue states correspond to innermost Majoranas at $\phi=0$. Starting from this point, varying $\phi$ causes the blue states to delocalize along the junction. A similar behavior is found on the green states with variations of $\mu_E$ outside the sweet spot. Changing $t_J$, however, does not cause any change in the wave functions of the sub--gap states. 

The fact that the eigenstates of the system have non--negligible values outside the low--energy subspace points to a limitation of the projection performed in the previous section, which is only valid close to the sweet spot.
As we discuss in what follows, a low--energy subspace that is written in terms of many--body occupations (even and odd) of the system is much more powerful. Starting first from the four Majoranas projection written in the many--body fermionic occupation basis (Appendix I.C), we obtain the corresponding subgap Josephson potential (Eq. 5 in the main text). In Appendix I.D, we go beyond this picture and describe the effective low--energy physics of the problem in terms of total many--body occupations including contributions from the four QDs (eight MBSs) forming the Josephson junction which allows us to obtain a subgap Josephson potential that includes terms containing both $\mu_E$ and $\mu_I$ on equal footing and to all orders (Eq. 8 of the main text).

\subsection{Projection in the left/right fermionic parity basis}
We can now write the matrix elements of $V_{\mathrm{DQD}}^{JJ}$ in the fermionic parity basis $\ket{n_L,n_R}$. For the total even parity state, the effective Josephson coupling reads
\begin{equation}
V_{\mathrm{DQD}}^{JJ} = \left(\begin{matrix}
\brakettt{00}{H_\gamma}{00} & \brakettt{00}{H_\gamma}{11}
\\
\brakettt{11}{H_\gamma}{00} & \brakettt{11}{H_\gamma}{11}
\end{matrix}\right) \;.
\end{equation}

Since the parity states are defined such that (similar for $c_R,c^\dagger_R$)
\begin{equation}
\begin{aligned}
c_L^\dagger \ket{n_L,n_R} = \sqrt{n_L+1}\ket{n_L+1,n_R} \quad &,\quad c_L \ket{n_L,n_R} = \sqrt{n_L}\ket{n_L-1,n_R} \;,
\\
\hat{n}_L \ket{n_L,n_R} = c_L^\dagger c_L & \ket{n_L,n_R} = n_L \ket{n_L,n_R} \;,
\end{aligned}
\end{equation}
and, attending to the decomposition  of these fermionic operators in Majorana operators (\ref{eq:supplemental/Majorana_decomposition_QD}), we can write the following operations,
\begin{equation}
\begin{aligned}
& i\gamma_{\alpha,1}^A\gamma_{\alpha,2}^B\ket{00/11} = (2\hat{n}_\alpha - 1)\ket{00/11} = -/+\ket{00/11}\;,
\\
& \gamma_{L,1}^A \gamma_{R,1}^A \ket{00/11} = (c_Lc_R + c_Lc_R^\dagger - c_Rc_L^\dagger - c_R^\dagger c_L^\dagger)\ket{00/11} = -/+\ket{11/00}\;,
\\
& i\gamma_{L,1}^A \gamma_{R,2}^B \ket{00/11} = (c_Lc_R - c_Lc_R^\dagger - c_Rc_L^\dagger + c_R^\dagger c_L^\dagger) \ket{00/11} = \ket{11/00}\;,
\\
& i \gamma_{L,2}^B \gamma_{R,1}^A \ket{00/11} = (c_Lc_R + c_Lc_R^\dagger - c_Rc_L^\dagger + c_R^\dagger c_L^\dagger) \ket{00/11} = \ket{11/00}\;,
\\
& \gamma_{L,2}^B \gamma_{R,2}^B \ket{00/11} = (-c_Lc_R + c_Lc_R^\dagger + c_Rc_L^\dagger + c_R^\dagger c_L^\dagger) \ket{00/11} = +/-\ket{11/00}\;.
\end{aligned}
\end{equation}

Therefore, the sub--gap contribution written in the even fermionic parity basis is
\begin{equation}
\begin{aligned}
\brakettt{00}{H_\gamma}{00} & = -(\lambda_{L1,L2} + \lambda_{R1,R2}) \;,
\\
\brakettt{11}{H_\gamma}{11} & = \lambda_{L1,L2} + \lambda_{R1,R2} \;,
\\
\brakettt{00}{H_\gamma}{11} & = i\lambda_{L1,R1} + \lambda_{L1,R2} + \lambda_{L2,R1} - i\lambda_{L2,R2} \;,
\\
\brakettt{11}{H_\gamma}{00} & = -i\lambda_{L1,R1} + \lambda_{L1,R2} + \lambda_{L2,R1} + i\lambda_{L2,R2} \;,
\end{aligned}
\end{equation}
where $\lambda_{\alpha\beta}$ are the matrix elements of (\ref{Effective Josephson4Majo}). Finally, the sub--gap Josephson potential takes the form
\begin{equation} \label{eq:supplemental/Majorana-transmon/Josephson_potential}
\begin{aligned}
& V_{\mathrm{DQD}}^{JJ}(\phi) = 
\\
& \frac{1}{2}\left(\begin{matrix}
\frac{2(t+\Delta)(t-\Delta) - (\mu_{L,1}\mu_{L,2} + \mu_{R,1}\mu_{R,2})}{t+\Delta} & t_J\left(1 - \frac{\mu_{L,1}\mu_{R,2}}{(t+\Delta)^2}\right)\cos\frac{\phi}{2} - it_J\frac{\mu_{L,1}-\mu_{R,2}}{t+\Delta}\sin\frac{\phi}{2}
\\
t_J\left(1 - \frac{\mu_{L,1}\mu_{R,2}}{(t+\Delta)^2}\right)\cos\frac{\phi}{2} + it_J\frac{\mu_{L,1}-\mu_{R,2}}{t+\Delta}\sin\frac{\phi}{2} & \frac{(\mu_{L,1}\mu_{L,2} + \mu_{R,1}\mu_{R,2}) - 2(t+\Delta)(t-\Delta)}{t+\Delta}
\end{matrix}\right) \;.
\end{aligned}
\end{equation}

Therefore, we can split this sub--gap effective potential in three different terms acting on a pseudospin parity space --Eq. (4) of the main text--,
\begin{equation}
\begin{aligned}
\label{Josephson4Majo}
    V_{\mathrm{DQD}}^{JJ}(\phi) &= E_M\cos\frac{\phi}{2}\sigma_x + E_M^S\sin\frac{\cos}{2}\sigma_y + \lambda\sigma_z \;,
    \\
    & E_M = \frac{t_J}{2}\left(1 - \frac{\mu_{L,1}\mu_{R,2}}{(t+\Delta)^2}\right) \;,
    \\
    & E_M^S = t_J\frac{\mu_{L,1}-\mu_{R,2}}{2(t+\Delta)} \;,
    \\
    & \lambda = \frac{2(t+\Delta)(t-\Delta) - (\mu_{L,1}\mu_{L,2} + \mu_{R,1}\mu_{R,2})}{2(t+\Delta)} \;.
\end{aligned}
\end{equation}

It is straightforward to see that, when restricting ourselves to the symmetric case  $\mu_{L,1}=\mu_{R,2}=\mu_E$ and $\mu_{L,2}=\mu_{R,1}=\mu_I$, the Josephson potential reduces to Eq. (5) of the main text.

\section{Beyond the four Majoranas projection: projection onto a full many--body parity basis}

A reasonable alternative treatment of the problem is to choose as our new fermionic parity subspace the two lowest--energy many--body eigenstates $\{|O_L^-,O_R^-\rangle,\,|E_L^-,E_R^-\rangle\}$ of both chains isolated from each other ($t_J=0$), where
\begin{equation}
H_\alpha = \left(\begin{matrix}
0 & 0 & 0 & \Delta_\alpha \\
0 & -\mu_{\alpha,1} & -t_\alpha & 0 \\
0 & -t_\alpha & -\mu_{\alpha,2} & 0 \\
\Delta_\alpha & 0 & 0 & -(\mu_{\alpha,1}+\mu_{\alpha,2})
\end{matrix}\right)\;,
\end{equation}
is the many--body Hamiltonian of one chain in the basis of occupation states $\{\ket{00},\,\ket{10},\,\ket{01},\,\ket{11}\}$. Defining $\mu_\alpha=(\mu_{\alpha,1}+\mu_{\alpha,2})/2$ and $\delta_\alpha=(\mu_{\alpha,1}-\mu_{\alpha,2})/2$, its eigenstates and eigenenergies are
\begin{equation}
\begin{aligned}
\ket{O_\alpha^-} = \left(0,\, \Psi_{\alpha,1}^A,\, \Psi_{\alpha,1}^B,\, 0\right)^T \propto \left(0,\, \frac{2\delta_\alpha + \epsilon_{\alpha O}^+ - \epsilon_{\alpha O}^-}{2t_\alpha},\, 1,\, 0\right)^T \quad &,\quad \epsilon_{\alpha O}^- = -\mu_\alpha - \sqrt{t_\alpha^2+\delta_\alpha^2}\;,
\\
\ket{O_\alpha^+} = \left(0,\, \Psi_{\alpha,2}^A,\, \Psi_{\alpha,2}^B,\, 0\right)^T \propto \left(0,\, \frac{2\delta_\alpha - \epsilon_{\alpha O}^+ + \epsilon_{\alpha O}^-}{2t_\alpha},\, 1,\, 0\right)^T \quad &,\quad \epsilon_{\alpha O}^+ = -\mu_\alpha + \sqrt{t_\alpha^2+\delta_\alpha^2}\;,
\\
\ket{E_\alpha^-} = \left(\Psi_{\alpha,3}^A,\, 0,\, 0,\, \Psi_{\alpha,3}^B\right)^T \propto \left(\frac{-\epsilon_{\alpha E}^+}{\Delta_\alpha},\, 0,\, 0,\, 1\right)^T \quad &,\quad \epsilon_{\alpha E}^- = -\mu_\alpha - \sqrt{\Delta_\alpha^2+\mu_\alpha^2}\;,
\\
\ket{E_\alpha^+} = \left(\Psi_{\alpha,4}^A,\, 0,\, 0,\, \Psi_{\alpha,4}^B\right)^T \propto \left(\frac{-\epsilon_{\alpha E}^-}{\Delta_\alpha},\, 0,\, 0,\, 1\right)^T \quad &,\quad \epsilon_{\alpha E}^+ = -\mu_\alpha + \sqrt{\Delta_\alpha^2+\mu_\alpha^2}\;.
\end{aligned}
\end{equation}

To construct the Hamiltonian of the junction living in the bipartite Hilbert space $\mathcal{H}_L\otimes\mathcal{H}_R$, we represent it on the basis of joint eigenstates $\{|i_L,j_R\rangle=|i_L\rangle\otimes|j_R\rangle\}$ with $i,j=O^\pm, E^\pm$. Thus, the Hamiltonian $\tilde{H}^{JJ}_{\mathrm{DQD}}=\tilde{H}_L+\tilde{H}_R+\tilde{H}_J$ has a diagonal term
\begin{equation}
\begin{aligned}
\tilde{H}_L+\tilde{H}_R &= (P_L^{-1} H_L P_L)\otimes\mathbb{I}_R + \mathbb{I}_L\otimes(P_R^{-1} H_R P_R) 
\\
&= \mathrm{diag}\left( \epsilon_{LO}^-,\,\epsilon_{LO}^+,\,\epsilon_{LE}^-,\,\epsilon_{LE}^+ \right) \otimes \mathbb{I}_R + \mathbb{I}_L\otimes \mathrm{diag}\left( \epsilon_{RO}^-,\,\epsilon_{RO}^+,\,\epsilon_{RE}^-,\,\epsilon_{RE}^+ \right) \;,
\end{aligned}
\end{equation}
where $P_\alpha$ is the change--of--basis matrix onto the eigenbasis of each chain. On the other hand, the off--diagonal term  $\tilde{H}_J$ is due to the Josephson tunneling between both chains, which can be easily represented on the joint--occupation basis $\{|n_{L,1},n_{L,2}\rangle\otimes|n_{R,1},n_{R,2}\rangle\}_{n_{\alpha,i}=0,1}$ and then projected onto the eigenbasis by the change--of--basis matrix $P_{LR}=P_L\otimes P_R$.

Finally, the Josephson potential (ignoring higher--order contributions from the rest of the eigenstates) can be written as
\begin{equation}
V^{JJ}_{\mathrm{DQD}} = \left(\begin{matrix}
\langle O_L^-,O_R^-| \tilde{H}^{JJ}_{\mathrm{DQD}} | O_L^-,O_R^-\rangle & \langle O_L^-,O_R^-|\tilde{H}^{JJ}_{\mathrm{DQD}}|E_L^-,E_R^-\rangle
\\
\langle E_L^-,E_R^-| \tilde{H}^{JJ}_{\mathrm{DQD}}|O_L^-,O_R^-\rangle & \langle E_L^-,E_R^-|\tilde{H}^{JJ}_{\mathrm{DQD}}|E_L^-,E_R^-\rangle
\end{matrix}\right) \;,
\end{equation}
where
\begin{equation}
\begin{aligned}
\langle O_L^-,O_R^-|\tilde{H}^{JJ}_{\mathrm{DQD}}|O_L^-,O_R^-\rangle &= \epsilon_{LO}^- + \epsilon_{RO}^-
\\
\langle E_L^-,E_R^-|\tilde{H}^{JJ}_{\mathrm{DQD}}|E_L^-,E_R^-\rangle &= \epsilon_{LE}^- + \epsilon_{RE}^-
\\
\langle E_L^-,E_R^-|\tilde{H}^{JJ}_{\mathrm{DQD}}|O_L^-,O_R^-\rangle &= t_J \left(4t^2 \sqrt{\frac{\epsilon_{RE}^+}{\epsilon_{LE}^+}} e^{i\phi/2} + \sqrt{\frac{\epsilon_{LE}^+}{\epsilon_{RE}^+}} (2\delta_L + \epsilon_{LO}^- - \epsilon_{LO}^+)(2\delta_R + \epsilon_{RO}^- - \epsilon_{RO}^+)e^{-i\phi/2}\right)
\\
& \times\frac{\Delta\sqrt{(2\delta_L + \epsilon_{LO}^+ - \epsilon_{LO}^-)(2\delta_R + \epsilon_{RO}^+ - \epsilon_{RO}^-)}}{8t^2\sqrt{(\epsilon_{LO}^+-\epsilon_{LO}^-)(\epsilon_{RO}^+-\epsilon_{RO}^-)(\epsilon_{LE}^+-\epsilon_{LE}^-)(\epsilon_{RE}^+-\epsilon_{RE}^-)}}
\\
& = -t_J \Psi_{L,1}^A\Psi_{R,1}^A \left( \Psi_{L,3}^B\Psi_{R,3}^A e^{i\phi/2} - \Psi_{L,4}^B\Psi_{R,4}^A \frac{\Psi_{L,2}^A\Psi_{R,2}^A}{\Psi_{L,2}^B\Psi_{R,2}^B} e^{-i\phi/2}\right) \;.
\end{aligned}
\end{equation}

One can see that, if the chemical potentials are constrained to the special symmetric choice $\mu_{L,1}=\mu_{R,2}=\mu_E$ and $\mu_{L,2}=\mu_{R,1}=\mu_E$ (internal vs. external), such that $\mu_L=\mu_R=\mu_E+\mu_I=\mu$ and $\delta_L=-\delta_R=\mu_E-\mu_I=\delta$, and considering $\Delta_L=\Delta_R$ and $t_L=t_R$, this Josephson potential reduces to the simpler form --Eq. (7) of the main text--
\begin{equation} \label{eq:supplemental/VJ_EO_IE}
V^{JJ}_{\mathrm{DQD}}(\phi)= \left(
\begin{matrix}
-2\mu - 2\sqrt{t^2+\delta^2} & \frac{t_J \Delta t}{2\sqrt{(t^2+\delta^2)(\Delta^2+\mu^2)}}\cos(\phi/2)
\\
\frac{t_J \Delta t}{2\sqrt{(t^2+\delta^2)(\Delta^2+\mu^2)}}\cos(\phi/2) & -2\mu - 2\sqrt{\Delta^2+\mu^2}
\end{matrix} \right) \;.
\end{equation}

\section{Majorana polarization}

The Hamiltonian $H^{JJ}$ described above can be separated into two independent blocks of even ($\{|O_L^\pm,O_R^\pm\rangle$, $|E_L^\pm,E_R^\pm\rangle\}$) and odd ($\{|E_L^\pm,O_R^\pm\rangle,\,|E_L^\pm,O_R^\pm\rangle\}$) total parity, which leads to a two--fold degenerate spectrum. To determine whether these degeneracies are associated with MBSs, we use the Majorana polarization (MP). This magnitude quantifies the MBS quality and is defined as the degree that a Hermitian operator localized on one of the quantum dots can switch between the lowest--energy states of even and odd blocks,
\begin{equation}
\begin{aligned}
& \mathrm{MP}_{\alpha,i}(O,E) =\frac{w_{\alpha,i}^2 - z_{\alpha,i}^2}{w_{\alpha,i}^2 + z_{\alpha,i}^2} \;,
\\
& w_{\alpha,i} = \brakettt{O}{c_{\alpha,i}+c_{\alpha,i}^\dagger}{E} \;,
\\
& z_{\alpha,i} = \brakettt{O}{c_{\alpha,i}-c_{\alpha,i}^\dagger}{E} \;.
\end{aligned}
\end{equation}

We can see that, for $t_J=0$, MP can be written as
\begin{equation}
\mathrm{MP}_{\alpha,i}= \frac{t_\alpha\Delta_\alpha}{(-1)^{i+1}\delta_\alpha\mu_\alpha - \sqrt{(t_\alpha^2+\delta_\alpha^2)(\Delta_\alpha^2+\mu_\alpha^2)}} \;,
\end{equation}
where $\ket{E}=|O_L^-,O_R^-\rangle$, $\ket{O}_{\alpha=L}=|E_L^-,O_R^-\rangle$, $\ket{O}_{\alpha=R}=|O_L^-,E_R^-\rangle$. Restricting ourselves to $t_\alpha=\Delta_\alpha$, $|\mathrm{MP}_{\alpha,1}|$ ($|\mathrm{MP}_{\alpha,2}|$) is maximum when $\mu_\alpha=\delta_\alpha$ ($\mu_\alpha=-\delta_\alpha$), that is, when $\mu_{\alpha,2}=0$ ($\mu_{\alpha,1}=0$).

Furthermore, from (\ref{eq:supplemental/VJ_EO_IE}), the effective Majorana coupling $E_M$ is related to this quantity such that
\begin{equation}
E_M = \frac{-t_J\mathrm{MP}_{\alpha,i}/2}{1 + (-1)^{i+\alpha}\frac{\delta\mu}{t\Delta}\mathrm{MP}_{\alpha,i}} \;,
\end{equation}
where $\alpha=\{0\equiv L,\, 1\equiv R\}$. Thus, if $\mu_E=\mu_I$ ($\mu_E=-\mu_I$), that is, $\delta=0$ ($\mu=0$), then $E_M$ is proportional to MP: $E_M=-t_J\mathrm{MP}/2$.

\section{Intraband splitting in transmon regime}

At $n_g=0.25$, the energy splitting between the ground state and the first excited state is merely due to the sub--gap Josephson potential since the rest of terms on the qubit Hamiltonian give rise to a doubly degenerate state at this point. Hence, it is reasonable to express the Kitmon qubit frequency $\omega_{KiT}\equiv\omega_{01}$ as the difference between the two eigenvalues of $V^{JJ}_{\mathrm{DQD}}(\phi)$,
\begin{equation}\label{eq:supplemental/VJ-eigenvalues}
    \Delta E^{JJ}(\phi) = 2 \sqrt{(\sqrt{t^2+\delta^2}-\sqrt{\Delta^2+\mu^2})^2 + E_M^2\cos^2\frac{\phi}{2}} \;.
\end{equation}

As we can see, this difference depends on $\phi$ and, hence, one should know the explicit form of the qubit wave functions to relate this quantity to $\omega_{01}$. Nevertheless, in the deep transmon regime ($E_J/E_C\gg 1$) these eigenfunctions can be approximated to harmonic--oscilator states sharpened around $\phi_{ext}$, so that the Kitmon frequency is $\omega_{KiT}\approx \Delta E^{JJ}(\phi_{ext})$ --Eq. (12) of the main text. Likewise, in transmon regime the qubit spectrum is insensitive to changes in the charge offset $n_g$, being this approximation valid for every parametric configuration of the system, even when diagonal terms of $V^{JJ}_{\mathrm{DQD}}(\phi)$ are not equal and these avoided crossings do not occur at $n_g=0.25$ in charging regime.

\begin{figure}
\centering
    \includegraphics[width=1\linewidth]{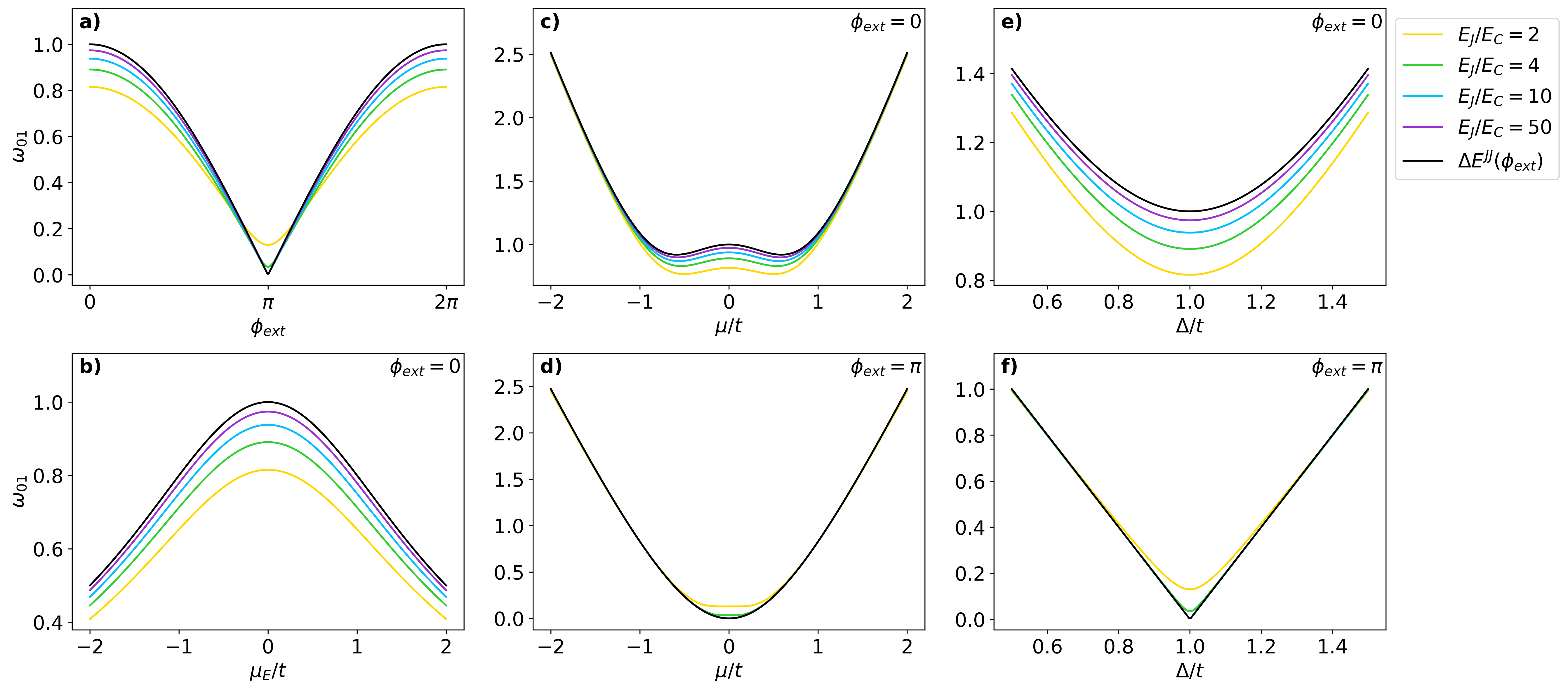}
\caption{Transition frequency $\omega_{01}$ for $E_J/E_C=2,4,10,50$, compared to analytical result (\ref{eq:supplemental/VJ-eigenvalues}), black line, as a function of (a) $\phi_{ext}$ at the sweet spot; (b) $\mu_E$ with $\mu_I=0$, $\Delta=t$ and $\phi_{ext}=0$; (c,d) $\mu_E=\mu_I=\mu$ with $\Delta=t$ and $\phi_{ext}=0,\pi$, respectively; and (e,f) $\Delta/t$ with $\mu_E=\mu_I=0$ and $\phi_{ext}=0,\pi$, respectively. We have fixed $t_J/t=1$ for all panels.}
\label{fig:supplemental/EJ-increasing}
\end{figure}

Fig. \ref{fig:supplemental/EJ-increasing} displays the transition frequency $\omega_{01}(n_g=0.25)$ as a function of different parameters, showing their evolution with increasing $E_J/E_C$ ratios. We show the convergence to $\Delta E^{JJ}(\phi_{ext})$ in the limit $E_J/E_C\gg 1$.

We can also check numerically this approximation by calculating the distance between the curves that the analytical result (\ref{eq:supplemental/VJ-eigenvalues}) and $\omega_{01}$ trace for increasing $E_J/E_C$ ratios. The distance between two curves described by the functions $f(x)$ and $g(x)$ over a parametric trajectory $x\in\mathcal{X}$ is written as
\begin{equation}
    d(f,g) = \left(\int_\mathcal{X} dx\,|f(x) - g(x)|^2\right)^{1/2} \;.
\end{equation}

\begin{figure}
\centering
    \includegraphics[width=0.35\linewidth]{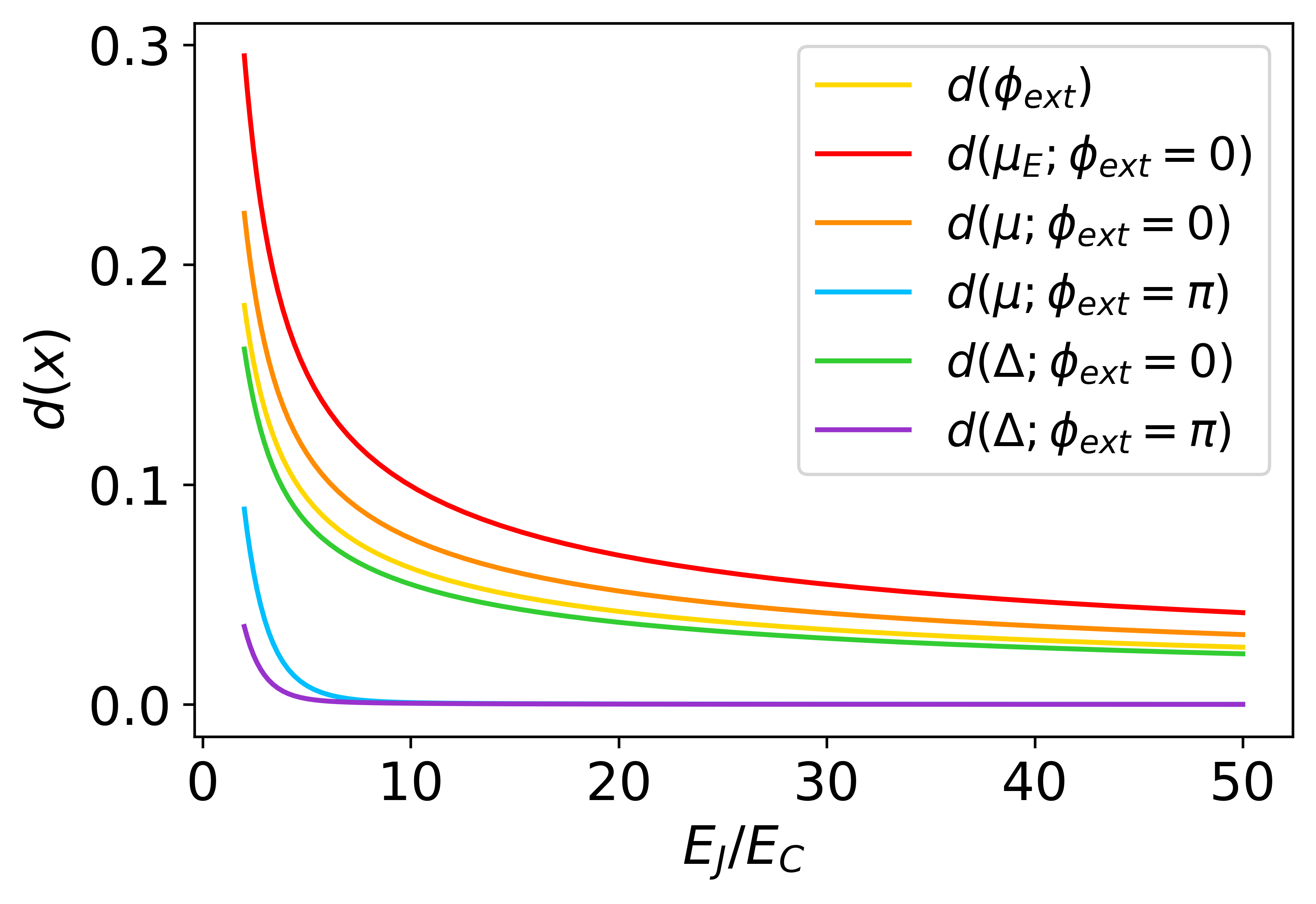}
\caption{Distance between curves $\omega_{01}$ and $\Delta E^{JJ}(\phi_{ext})$ as a function of $E_J/E_C$ for the same curves shown in Fig. \ref{fig:supplemental/EJ-increasing} (see legend).}
\label{fig:supplemental/curve-distance}
\end{figure}

As we can observe in Fig. \ref{fig:supplemental/curve-distance}, increasing the ratio $E_J/E_C$ minimizes the distance between numerical results and our analytical approximation, which allows us to predict $\omega_{KiT}$ with great precision in the deep transmon regimen.

\begin{figure}[ht]
    \centering
    \includegraphics[width=\linewidth]{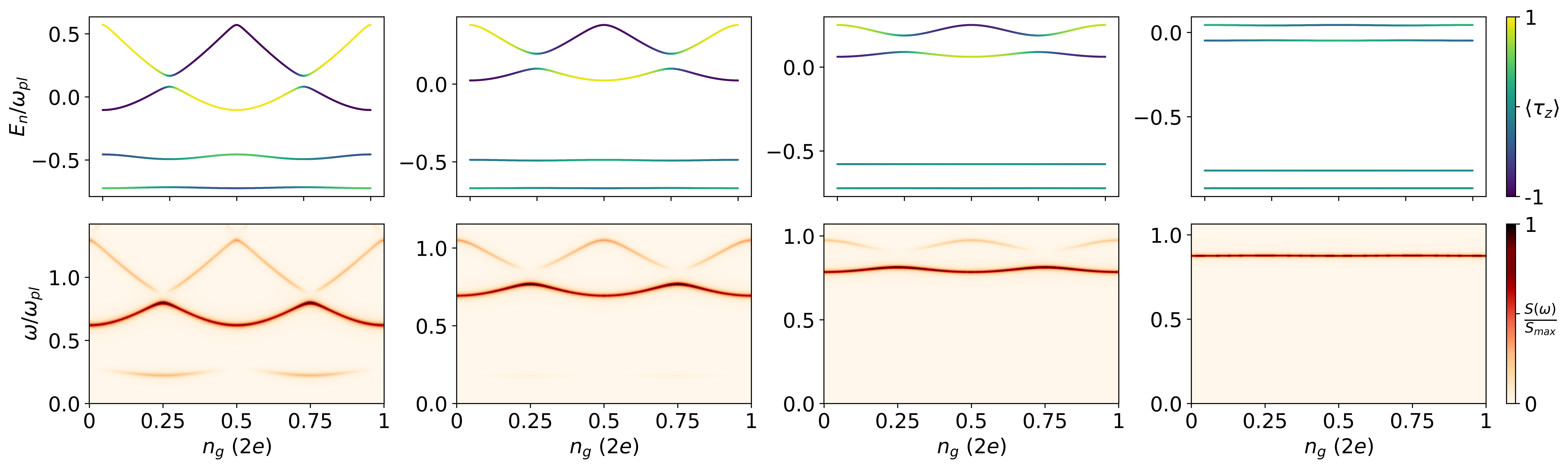}
    \caption{Full evolution of the energy spectrum and its MW response as a function of $n_g$ at the sweet spot ($\phi_{ext}=0$) for $E_J/E_C=1.5, 3, 5, 10$ (from left to right).}
    \label{fig:supplemental/supp_panel-ng}
\end{figure}

\begin{figure}[ht]
    \centering
    \includegraphics[width=\linewidth]{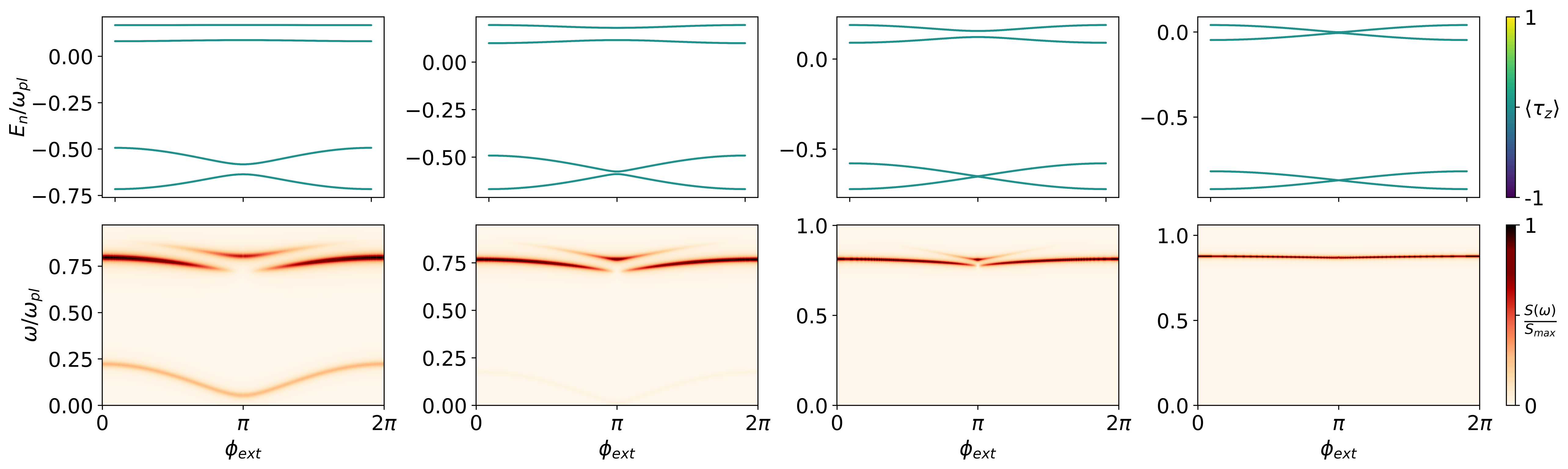}
    \caption{Full evolution of the energy spectrum and its MW response as a function of $\phi_{ext}$ at the sweet spot for $E_J/E_C=1.5, 3, 5, 10$ (from left to right).}
    \label{fig:supplemental/supp_panel-phi}
\end{figure}

Finally, we include some additional results that show a full progression of the energy spectrum and its MW response for increasing $E_J/E_C$ ratios. In particular, we can see in Fig. \ref{fig:supplemental/supp_panel-ng} an enhancement of the insensitivity to the charge offset as the qubit enters in the transmon regime, with a dominant transition $\omega_{02}$. Furthermore, Fig. \ref{fig:supplemental/supp_panel-phi} shows how the spectral hole in $\omega_{02}$ at $\phi_{ext}$ narrows until true energy crossing appears as the $E_J/E_C$ ratio increases.

\section{Numerical methods for the Majorana--transmon qubit: tight--binding treatment} \label{sec:supplemental/Majorana-transmon/tight-binding}

\subsection{Phase space}

In phase space, the numerical solution of the qubit Hamiltonian
\begin{equation} \label{eq:supplemental/qubit_Hamiltonian}
    H_Q = 4E_C(\hat{n} - n_g)^2 + V_J(\phi)\;,
\end{equation}
is accomplished by discretizing the phase space as $\phi_j = 2\pi j/l^\phi$, with $j = 1, \dots, l^\phi$, defining a set of sites arranged into a circular chain. In so doing, the Hamiltonian acquires a tight--binding form and it allows us to define a finite fermionic Hilbert space and operators $b_j^{(\dagger)}$ such that their action on the ground state is $b_j^\dagger\ket{0}=\Psi(\phi_j)$, where $\Psi(\phi)$ is the eigenstate at phase $\phi$.

Then, starting from the definition of the derivative
\begin{equation}
\frac{df(x)}{dx} = \lim_{h\to 0} \frac{f(x+h) - f(x-h)}{2h} \;,
\end{equation}
we can express the operator $\hat{n}=-i\partial_\phi$ in the discretized form
\begin{equation}
-i\partial_\phi = -i\frac{(b_{i+1}^\dagger - b_{i-1}^\dagger)b_i}{2a_\phi} \;,
\end{equation}
where $a_\phi = 2\sin(\pi/l^\phi)$ is a phase lattice constant. By construction, the second derivative is defined as
\begin{equation}
\frac{d^2f(x)}{dx^2} = \lim_{h\to 0} \frac{f(x+h)-2f(x)+f(x-h)}{h^2} \;,
\end{equation}
so we can write
\begin{equation}
\partial^2_\phi = \frac{(b_{i+1}^\dagger - 2b_i^\dagger +b_{i-1}^\dagger)b_i}{a_\phi^2} \;.
\end{equation}

Hence, the Hamiltonian (\ref{eq:supplemental/qubit_Hamiltonian}) reads
\begin{equation}
\begin{aligned}
H & = \sum_j b_j^\dagger h_j^\phi b_j + \sum_{\langle j,k\rangle} b_j^\dagger v_{jk}^\phi b_k \;,
\\
& h_j^\phi = 4E_C(2a_\phi^{-2} + n_g^2) + V_J(\phi_j) \;,
\\
& v_{jk}^\phi = 4E_C[\sgn(j-k) i n_g a_\phi^{-1} - a_\phi^{-2}] \;,
\end{aligned}
\end{equation}
where each site element $h_j^\phi,v_{jk}^\phi$ is a $2\times 2$ matrix, owing to the pseudospin structure from even--odd projection.

Secondly, the eigenstates of the Hamiltonian (\ref{eq:supplemental/qubit_Hamiltonian}) are defined as a two--component spinor $\Psi_k=(f_k(\phi),g_k(\phi))^T$ with periodic/antiperiodic boundary conditions in phase space, $f(\phi+2\pi)=f(\phi)$ and $g(\phi+2\pi)=-g(\phi)$, due to their even/odd fermionic parity. To make the Hamiltonian fully periodic, it is rotated according to $H(\phi)\to UH(\phi)U^\dagger$, with $U=\diag(1,e^{i\phi/2})$. Therefore, the final form of the Hamiltonian (\ref{eq:supplemental/qubit_Hamiltonian}) is
\begin{equation}
H = \left(\begin{matrix}
h(n_g) + V_J^{11} & V_J^{12} e^{-i\frac{\phi}{2}} \\ e^{i\frac{\phi}{2}} V_J^{21} & h\left(n_g +\frac{1}{2}\right) + e^{i\frac{\phi}{2}} V_J^{22} e^{-i\frac{\phi}{2}}
\end{matrix}\right)\;,
\end{equation}
and hence the site elements $h_j^\phi$ and $v_{jk}^\phi$ change according to this transformation.

\subsection{Charge space}

In charge representation, the set of states $\{\ket{n}\}_{n=-\infty}^{\infty}$ form a orthonormal basis of such space. Here, the number of Cooper pairs operator is defined as
\begin{equation}
\hat{n} = \sum_{n=-\infty}^\infty n \ket{n}\bra{n} \;,
\end{equation}
whereas the action of its conjugate operator $\phi$ on each one of these states is
\begin{equation}
e^{ik\phi}\ket{n} = \ket{n+k} \;.
\end{equation}

Therefore, the Hamiltonian (\ref{eq:supplemental/qubit_Hamiltonian}) can be expressed as
\begin{equation}
H = \sum_{n=-\infty}^\infty (n-n_g)^2 \ket{n}\bra{n} + V_J(\phi) \;,
\end{equation}
where the form of the Josephson potential is conditioned by its phase--dependent terms, being
\begin{equation}
\begin{aligned}
\cos(k\phi) & = \frac{1}{2}\sum_{n=-\infty}^\infty \left(\ket{n+k}\bra{n} + \hc\right)\;,
\\
\sin(k\phi) & = \frac{-i}{2}\sum_{n=-\infty}^\infty \left(\ket{n+k}\bra{n} - \hc\right)\;,
\end{aligned}
\end{equation}
the most usual of them. Indeed, for more complex potentials, we can perform a Fourier transform which reduces it to a simple sum of these terms. This representation gives rise to an identical spectrum to that calculated in phase space. However, in this case, we require a smaller (truncated) number of sites $N$ of the tight--binding Hamiltonian matrix, so this method needs less computational power and time than the other one. Note that, in phase space, $\dim=2N$ since each site is a spinor with two possible parities, whereas in charge space we have a set of states $\{\ket{n}\}$ ($n=-N,-N+1/2, \dots, 0, 1/2, \dots, N$, so that $\dim=2N+1$.

\begin{figure}
\centering
    \includegraphics[width=0.8\linewidth]{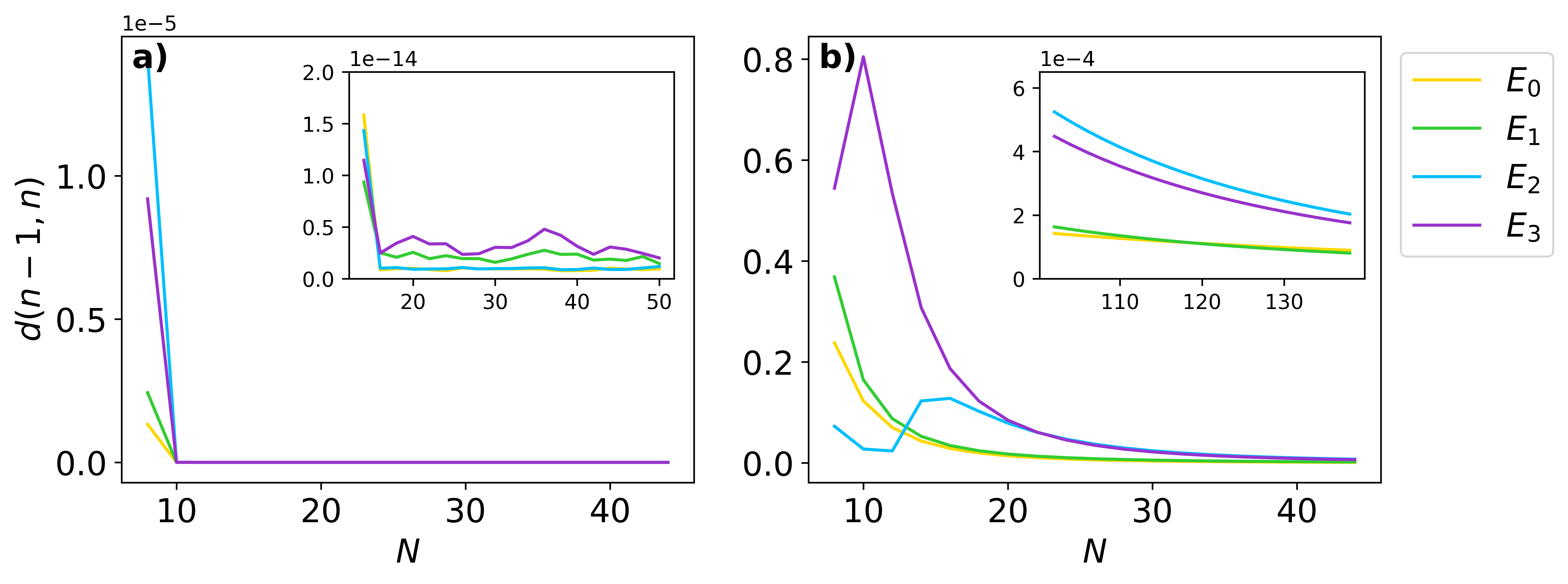}
\caption{Distance between curves $E_i^{N-1}(n_g)$ and $E_i^{N}(n_g)$ (where $i=0,1,2,3$ labels eigenstates of increasing energy) at the sweet spot as a function of a cutoff $N$. Numerical methods are implemented in (a) charge space and (b) phase space.}
\label{fig:supplemental/convergence}
\end{figure}

Indeed, Fig. \ref{fig:supplemental/convergence} shows the convergence of the first four states as a function of $N$, defined as the maximum number of sites that discretize the tight--binding space. This convergence is defined as the distance between the curves that each eigenstate traces (as a function of $n_g$) with $N-1$ and $N$ sites. It is straightforward to see that the tight--binding method converges much faster in charge space than in phase space.

\end{document}

%% file: preamble.tex
\usepackage{lmodern}
\usepackage[T1]{fontenc}
\usepackage[english]{babel}
\usepackage{graphicx}
\usepackage[caption=false]{subfig}
\usepackage{amsmath}
\usepackage{mathrsfs}
\usepackage{amsfonts}
\usepackage{amsthm}
\usepackage{amssymb}
\usepackage{enumerate}
\usepackage{wrapfig}
\usepackage{mathtools}
\usepackage[colorlinks=true,linkcolor=blue]{hyperref}
\usepackage{xcolor}
\usepackage{physics}
\usepackage{listings}

\definecolor{codegreen}{rgb}{0,0.6,0}
\definecolor{codegray}{rgb}{0.5,0.5,0.5}
\definecolor{codepurple}{rgb}{0.58,0,0.82}
\definecolor{backcolour}{rgb}{0.95,0.95,0.92}

\lstdefinestyle{mystyle}{
    backgroundcolor=\color{backcolour},   
    commentstyle=\color{codegreen},
    keywordstyle=\color{magenta},
    numberstyle=\tiny\color{codegray},
    stringstyle=\color{codepurple},
    basicstyle=\ttfamily\footnotesize,
    breakatwhitespace=false,         
    breaklines=true,                 
    captionpos=b,                    
    keepspaces=true,                 
    showspaces=false,                
    showstringspaces=false,
    showtabs=false,                  
    tabsize=2
}

\newtheorem{teo*}{Teorema}
	
\newtheorem{lem*}{Lema}					

\newtheorem{deff*}{Definición}

\newtheorem{prop*}{Proposición}

\theoremstyle{definition}

\DeclareMathOperator{\sgn}{sgn}

\DeclareMathOperator{\hc}{h.c.}

\usepackage{stackengine}
\stackMath

\newcommand{\brakettt}[3] {{
	\left\langle #1 \right| #2 \left| #3 \right\rangle
}}
\newcommand{\diag}{{
	\text{diag}\,
}}